\documentclass[a4paper,onecolumn,11pt,accepted=2025-01-08]{quantumarticle}
\pdfoutput=1
\usepackage[utf8]{inputenc}
\usepackage[english]{babel}
\usepackage[T1]{fontenc}
\usepackage{lipsum}
\usepackage{bbm}

\usepackage{braket}
\usepackage{amsfonts}
\usepackage{amssymb}
\usepackage{amsmath}
\usepackage{latexsym}
\usepackage{amsthm}
\usepackage[usenames]{color}
\usepackage{hyperref}
\usepackage{optidef}
\usepackage[capitalise]{cleveref}
\usepackage[parfill]{parskip}
\usepackage{bm}
\usepackage{tikz}
\usepackage{caption}
\theoremstyle{definition}

\hypersetup{pdfpagemode=UseNone}
\newtheorem{protocol}{Protocol}
\newtheorem{theorem}{Theorem}[section]
\newtheorem{lemma}[theorem]{Lemma}

\theoremstyle{definition}

\usepackage{color,graphicx}

\newcommand{\A}{\mathcal{A}}
\newcommand{\B}{\mathcal{B}}
\newcommand{\Y}{\mathcal{Y}}

\newcommand{\inner}[2]{\langle #1, #2 \rangle}
\newcommand{\ketbra}[2]{\ket{#1}\bra{#2}} 
 
\newcommand{\kb}[1]{\ket{#1}\bra{#1}} 
\newcommand{\Tr}{\mathrm{Tr}} 
\newcommand{\D}{\textup{D}}
\newcommand{\task}{\mathrm{Task}}

\begin{document}

\title{Breaking barriers in two-party quantum cryptography via stochastic semidefinite programming}

\author{Akshay Bansal}
\affiliation{Department of Computer Science, Virginia Polytechnic Institute and State University, Blacksburg, Virginia, USA}
\email{akshaybansal14@gmail.com}

\author{Jamie Sikora}
\affiliation{Department of Computer Science, Virginia Polytechnic Institute and State University, Blacksburg, Virginia, USA}
\email{sikora@vt.edu}
\homepage{https://sites.google.com/site/jamiesikora/}

\maketitle

\begin{abstract}
In the last two decades, there has been much effort in finding secure protocols for two-party cryptographic tasks. 
It has since been discovered that even with quantum mechanics, many such protocols are limited in their security promises. 
In this work, we use stochastic selection, an idea from stochastic programming, to circumvent such limitations. For example, we find a way to \emph{switch} between bit commitment, weak coin flipping, and oblivious transfer protocols to improve their security. 
We also use stochastic selection to turn trash into treasure yielding the first quantum protocol for Rabin oblivious transfer. 
\end{abstract}


\section{Introduction} 

The first work in the quantum cryptography was by Wiesner in~\cite{wiesner1983conjugate}. 
In that seminal paper, he introduced two cryptographic tasks which are still very much studied by the quantum community, quantum money and a task which he called \emph{multiplexing}. 
Multiplexing is now better known as $1$-out-of-$2$ oblivious transfer in which Bob wishes to send to Alice one bit while keeping the other bit hidden. 
Wiesner's protocol was knowingly insecure in the unconditional security model, where no computational bounds are put on the adversaries, but he argued that it would be infeasible to break using (then) current technology (which is an argument which could still be made to this day, some $40$ years later). 

The term \emph{oblivious transfer} was first introduced by Rabin in 1981~(see \cite{rabin2005exchange}).
The task studied in that paper was Alice sending to Bob a single bit which he receives with probability $1/2$ and is lost with probability $1/2$. 
This task which is sometimes now referred to Rabin oblivious transfer has been showed to be equivalent to $1$-out-of-$2$ oblivious transfer (for some definition of equivalent) by Cr\'{e}peau in 1994~\cite{crepeau1994quantum}. 
In this work, we shall refer to multiplexing/$1$-out-of-$2$ oblivious transfer simply as \emph{oblivious transfer} and Rabin's similar task as \emph{Rabin oblivious transfer}. 

In addition to these tasks, there are several other two-party cryptographic tasks studied in this work. 
We briefly introduce them below, but we note now a common story; that most of them are provably insecure by a significant margin. 
Indeed, there are general results proving the insecurity of many such cryptographic tasks, see for example~\cite{osborn2022constant, schaffner2007cryptography}. 
This brings us to the main question which motivates this work. 
\begin{equation*} 
    \emph{Can we do anything about the limited unconditional security of quantum two-party cryptography?}
\end{equation*} 
To better understand these limitations, we now give a glimpse into the history of quantum protocols for two-party cryptography (with rigorous definitions deferred to more formal discussions later). 

\subsection{A brief history of two-party cryptography} 

We now give a brief history of several cryptographic tasks and in doing so, indicate bounds on their security. 
When relevant, we also discuss further limitations when considering \emph{simple} protocols. 

\paragraph{Oblivious transfer.} 
Several constant lower bounds on oblivious transfer have been presented in~\cite{chailloux2013lowerot, buhrman2012complete,osborn2022constant,amiri2021imperfect} with the largest being in~\cite{chailloux2016semihonest}. 
In that work, it was shown that Alice or Bob can cheat with probability at least $2/3$, a large gap above the perfect security bound of $1/2$. 
The best known explicit protocol\footnote{Using slightly different security definitions, the protocol in~\cite{amiri2021imperfect} has better security parameters concerning one of the cheating parties.} that we have is from~\cite{chailloux2013lowerot} which attains cheating probabilities both equal to $3/4$. 
(Definitions of cheating probabilities are deferred to  formal discussions later.)
 
We also discuss Rabin oblivious transfer in this paper. 
As far as we are aware, there are no quantum protocols known for Rabin oblivious transfer. 
However, there is a constant lower bound known for a particular variant of it~\cite{osborn2022constant}. 

\paragraph{Bit commitment.} 
Bit commitment is the task where Alice commits to a bit $y$, then wishes to reveal it at a later time. 
Cheating Alice wants to be able to reveal any bit she chooses later and a cheating Bob wants to learn $y$ before it is revealed. 
It has been showed that Alice or Bob can cheat with probability at least $0.739$. 
Simple protocols have since been introduced where Alice and Bob can both cheat with probability $3/4$ with a matching lower bound for such protocols in~\cite{ambainis2001new}. 
Since we wish to have the cheating probabilities close to $1/2$, there is a significant security gap here as well.

\paragraph{Weak coin flipping.} 
Weak coin flipping is the task where Alice and Bob use a communication channel to generate a random bit but Alice and Bob want different outcomes. 
Mochon proved that it is possible to find arbitrarily good quantum protocols for this task~\cite{mochon2007quantum} (see also~\cite{aharonov2016simpler}) which have been made explicit in~\cite{arora2019quantum, arora2021analytic}. 
However, these are quite complicated and it has since been shown that any such protocol requires a huge communication cost~\cite{miller2020impossibility}. 
Mochon also gave a fairly simple protocol called Dip-dip-boom, see~\cite{mochon2005large}, where Alice and Bob cannot influence the outcome with probability greater than $2/3$. 
This is, again, a constant above the preferred cheating probabilities of $1/2$, but it has the benefit of being relatively simple. 

\paragraph{Strong coin flipping and die rolling.}
Strong coin flipping is the same as weak coin flipping but Alice or Bob may try to influence the outcome towards either $0$ or $1$ (i.e., we do not assume they want different outcomes). 
Die rolling is the generalization of strong coin flipping to $D$ possible outcomes (strong coin flipping being the case when $D=2$).   
Kitaev used semidefinite programming~\cite{kitaev2002quantum} to prove that Alice or Bob can cheat with probability at least $1/\sqrt{2}$ in any quantum protocol for strong coin flipping which is easily extended to $1/\sqrt{D}$ for the general case of die rolling (see~\cite{aharon2010quantum} for details).  
In each case, the lower bound is a constant above the desired bound of $1/D$.  
  
\subsection{Our stochastic switching framework} \label{subsection:intro:stochasticSwitching}

We now illustrate our idea to improve the security of cryptographic tasks. 
Suppose we have two communication tasks between Alice and Bob (or even more parties) such that the first several rounds of communication are identical. 
These two protocols may end up diverging from one another and ultimately result in accomplishing two completely different goals. 
What we do is have Alice and Bob start the communication, then at some point, one of the parties (selected beforehand, say Bob) flips a coin and announces whether to continue with either the first task or the second. 
Then they agree to continue to complete the announced task. 

To take an example, suppose one task is oblivious transfer and the other task is bit commitment. 
They then communicate and at some pre-agreed upon point  Bob announces ``Alice, we are finishing this protocol as a bit commitment protocol'', to which  
Alice agrees. 

\paragraph{Is this a realistic setting?} 
Notice that in the above example, even though Alice and Bob end up doing bit commitment, this may not have been the case. 
Moreover, this protocol does not adhere to the traditional definition of a proper bit commitment protocol since there is a chance that it accomplishes something else entirely (oblivious transfer). 
In a way, this suggests a protocol for \emph{bit commitment or oblivious transfer} decided randomly from within the protocol itself. 
However, we argue that this is a realistic setting. 
In most scenarios, Alice and Bob are not set up to communicate in order to accomplish a single task a single time, but rather many varied tasks in succession. 
In this work, we study basic cryptographic tasks, sometimes referred to as primitives, which are used as building blocks in much more elaborate, more complicated protocols. 
Thus, it is entirely likely that Alice and Bob do not want to perform a single bit commitment protocol or a single weak coin flipping protocol, but rather perform them hundreds or thousands of times.   

Moreover, our stochastic switch framework is so versatile that it can be used to switch between many tasks, cryptographic or otherwise. 
Many of our tasks start with Alice creating the EPR state 
\begin{equation} 
\frac{1}{\sqrt 2}\ket{00} + \frac{1}{\sqrt 2}\ket{11}  
\end{equation} 
and sending half to Bob. 
This is reminiscent of many other basic tasks in quantum computing such as superdense coding, teleportation, and entanglement-swapping, to list a few. 
These do not a priori have anything to do with cryptography, but are communication primitives in their own right. 
Therefore, many of our examples in this paper can be used in a setting more general than just two-party cryptography; they can be used in general communication tasks. 
One setting to which our framework could easily be applied is a node-to-node communication point in a quantum internet where the two nodes may need to sometimes generate random numbers (weak coin flipping), retrieve information from a database (oblivious transfer), or use it as a quantum repeater to aid in quantum key distribution (entanglement-swapping). 

\paragraph{A familiar example where the stochastic switch is useful.} 
We now illustrate our switching idea in a simple, yet familiar, communication task. 

\begin{protocol}[A seemingly pointless protocol] \label{protocol:broadcasting:first}
\quad 
\begin{itemize} 
\item Suppose Alice creates a qubit $\ket{\psi}$ (unknown to Bob). 
She sends it to Bob who then returns a two-qubit state $\ket{\phi}$ to Alice with the intention that the first qubit of $\ket{\phi}$ is $\ket{\psi}$. 
\item Alice checks if the first qubit is $\ket{\psi}$ with the POVM: 
\begin{equation}
\{ \kb{\psi} \otimes \mathbbm{1}_2, \mathbbm{1}_4 - ( \kb{\psi} \otimes \mathbbm{1}_2) \}. 
\end{equation}
\end{itemize} 
\end{protocol} 
We note that Bob can easily pass Alice's test by simply returning $\ket{\psi} \otimes \ket{0}$.\\

\begin{protocol}[Another seemingly pointless protocol] \label{protocol:broadcasting:second}
\quad \\ 
The same as \cref{protocol:broadcasting:first}, except Alice checks if the second qubit is $\ket{\psi}$ instead. 
\end{protocol} 

Again, Bob can easily pass Alice's test, this time by returning $\ket{0} \otimes \ket{\psi}$. 

While these two protocols each allow Bob to pass the test, we now use a stochastic switch to make the game harder for Bob. 

\begin{protocol}[Switching between Protocols~\ref{protocol:broadcasting:first} and~\ref{protocol:broadcasting:second}] 
\quad 
\begin{itemize} 
\item Suppose Alice creates a qubit $\ket{\psi}$ (unknown to Bob). 
She sends it to Bob who then returns a two-qubit state $\ket{\phi}$. 
\item \textit{Alice selects $c\in\{0,1\}$ uniformly at random and sends it to Bob}. 
\item If $c = 0$, Alice checks to see if the \emph{first} qubit is in state $\ket{\psi}$. 
If $c = 1$, Alice checks to see if the \emph{second} qubit is in state $\ket{\psi}$. 
\end{itemize} 
\end{protocol} 

For Bob to pass this test, he must somehow make the first qubit equal to $\ket{\psi}$ and also the second qubit. 
Since this qubit is unknown to Bob it is impossible to win this game with certainty by the \emph{no-broadcasting theorem}~(see, e.g.,~\cite{barnum2007generalized}) which is a refinement of the no-cloning theorem~\cite{wootters2009no}. 
We could calculate the maximum probability with which  Bob could pass the test (if more details about the states are given), but since it must be strictly less than $1$, we have illustrated our point; the switch decreases Bob's probability to pass this test. 

\paragraph{Why would this improve security, and does it always?}  
A moment's thought reveals why switching between Protocols~\ref{protocol:broadcasting:first} and~\ref{protocol:broadcasting:second} ruins Bob's chances of passing Alice's test. 
The message Bob would send back to Alice in \cref{protocol:broadcasting:first} is different than the state Bob would send back to Alice in \cref{protocol:broadcasting:second}. 
Since the stochastic switch (Alice's selection of $c \in \{ 0,1 \}$) occurs \emph{after} Bob's returned message, he cannot simultaneously use both strategies. 
Thus, this limits what Bob can do. 

In fact, a specialized variant of stochastic switch also finds its application in interactive proof systems where an honest \emph{verifier} wishes to verify the zero-knowledge(ness) of a \emph{prover} by randomly picking a challenge (from a predefined set of challenges) independent of the prior interaction with the prover. This switching method can ensure the zero-knowledge behaviour of the \emph{prover}. In our work, we generalize this framework of stochastic switching by also allowing a selection amongst fundamentally different tasks and analyzing the statistical security by formulating it as stochastic semidefinite programming problem (discussed in detail in the later sections). Although the formulation and the ideas remain general, we mostly focus on quantum protocols where a natural analysis follows via semidefinite programming. It is worth noting that some of the more well-known techniques in classical cryptography, e.g., the \emph{cut-and-choose} methods and their application in secure two-party computation~\cite{lindell2016fast} also falls conceptually within our framework.    

\begin{figure}[hbt!]
    \tikzset{every picture/.style={line width=0.75pt}} 
    \centering
    \begin{tikzpicture}[x=0.75pt,y=0.75pt,yscale=-1,xscale=1]
    
    \draw   (223,166.25) .. controls (223,143.47) and (248.74,125) .. (280.5,125) .. controls (312.26,125) and (338,143.47) .. (338,166.25) .. controls (338,189.03) and (312.26,207.5) .. (280.5,207.5) .. controls (248.74,207.5) and (223,189.03) .. (223,166.25) -- cycle ;
    \draw   (372,166.25) .. controls (372,143.47) and (397.74,125) .. (429.5,125) .. controls (461.26,125) and (487,143.47) .. (487,166.25) .. controls (487,189.03) and (461.26,207.5) .. (429.5,207.5) .. controls (397.74,207.5) and (372,189.03) .. (372,166.25) -- cycle ;
    \draw   (275.43,248.35) .. controls (283.44,235.61) and (325.57,247.65) .. (369.55,275.26) .. controls (413.52,302.87) and (442.68,335.58) .. (434.68,348.33) .. controls (426.68,361.07) and (384.54,349.03) .. (340.57,321.42) .. controls (296.59,293.81) and (267.43,261.1) .. (275.43,248.35) -- cycle ;
    \draw   (276.75,350.37) .. controls (269.57,337.07) and (298.81,303) .. (342.05,274.26) .. controls (385.3,245.53) and (426.18,233.02) .. (433.36,246.31) .. controls (440.54,259.61) and (411.31,293.68) .. (368.06,322.42) .. controls (324.81,351.15) and (283.93,363.66) .. (276.75,350.37) -- cycle ;
    
    \draw (271,160.4) node [anchor=north west][inner sep=0.75pt]    {$S_{1}$};
    \draw (420,160.4) node [anchor=north west][inner sep=0.75pt]    {$S_{2}$};
    \draw (287,251.55) node [anchor=north west][inner sep=0.75pt]    {$S_{3}$};
    \draw (407,251.55) node [anchor=north west][inner sep=0.75pt]    {$S_{4}$};

    \end{tikzpicture}
    
\caption{Let $S_j$ denote the set of optimal first messages Alice could send in $\task_j$. Then if two sets are disjoint, Alice would have to hedge her cheating attempts if Bob switches between those two tasks (e.g., $\task_1$ and $\task_2$). 
If the two sets have a nonempty intersection, Alice would have no difficulty cheating if Bob switches between those two tasks (e.g., $\task_3$ and $\task_4$).}  \label{fig:optimalSets}   
\end{figure}
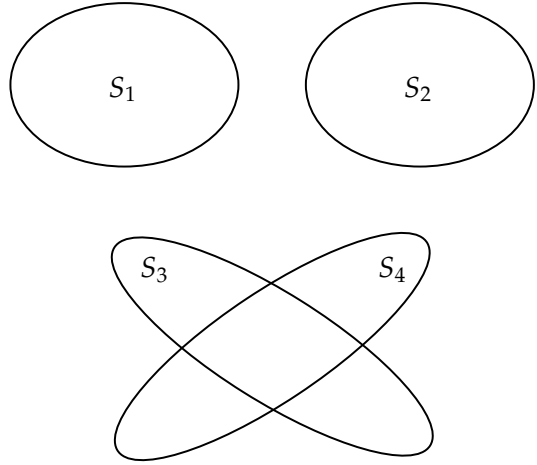

We now discuss this idea further and pin down, conceptually, when this may work and when it may not. 
For simplicity, suppose Alice sends a single message to Bob who then does the stochastic switch between two tasks, call them 
$\task_1$ and $\task_2$. 
Note that the first message must be modeled by the same density operator for them to be considered for this framework. 
However, Alice might be tempted to digress from this ``honest'' message in each task. 
Suppose for $\task_1$ Alice's optimal strategy involves her sending a first message modeled by the density operator $\sigma_1$ and, similarly, Alice's optimal strategy for $\task_2$ involves her sending a first message modeled by the density operator $\sigma_2$. 
Then if $\sigma_1 \neq \sigma_2$, then the two tasks could be good candidates to switch between to see a decrease in Alice's cheating. 
The story is a bit more nuanced than this though. 
Alice could have multiple choices for her first message in an optimal strategy. 
Therefore, we must consider the \emph{set of optimal first messages} in each scenario. 
\cref{fig:optimalSets} illustrates two scenarios.

As the above figure suggests, if the two sets of optimal first messages have no intersection, then Alice's cheating attempts are strictly decreased in this setting. 
A similar story can be said when considering switching between three or more protocols as well. 
 
If this setting is to improve security, we would require Bob to not be able to break the protocol by abusing his power to control the stochastic switch. 
We now describe the general model of a cheating Alice and cheating Bob (and give explicit formulations when analyzing specific instances in later sections). 

\paragraph{Modeling cheating Alice and cheating Bob.} 
We discuss the following notions of security for a \emph{complete} switch protocol with multiple possible subtasks. 
\begin{itemize}
\item \emph{Completeness:} A switch protocol is said to be (conditionally) \emph{complete} if whenever Alice and Bob are honest, then the protocol resulting from Bob's selection is \emph{complete}. 
In other words, if Bob chooses $\task_j$, the resulting protocol does exactly what $\task_j$ is supposed to do. 
    
\item \emph{Cheating Alice:} 
A cheating Alice could in general try to find a strategy which can accommodate for a good cheating attempt for every task choice of Bob. 
The cheating probability for dishonest Alice against honest Bob is given by 
    \begin{equation}\label{mathdef:cheatingAlice:first}
        P_A = \max \left\{ \sum_j p_j \; P_{A}^{\task_j} \right\} 
    \end{equation}
where the maximum is taken over all possible strategies of Alice, 
$p_j$ is the probability honest Bob chooses $\task_j$, and $P_A^{\task_j}$ is the maximum cheating probability for Alice conditioned on Bob choosing $\task_j$ in the context of the switching protocol.  

\item \emph{Cheating Bob:} 
Bob's cheating attempts could involve using his protocol switching choice to his advantage. 
As an example, if one task allows Bob to cheat with a large probability, he has the power to simply choose that task. 
A slightly more sophisticated approach is for him to perform a measurement and then, conditioned on the outcome, make his choice on how to proceed. 
    
The cheating probability for dishonest Bob against honest Alice is given by 
\begin{equation}\label{mathdef:cheatingBob:first}
    P_B = \max \left\{ \sum_j \Pr[\text{Bob chooses $\task_j$}]P_{B}^{\task_j} \right\} 
\end{equation}
where the maximum is taken over all strategies of Bob and $P_{B}^{\task_j}$ is the maximum cheating probability for Bob conditioned on him choosing $\task_j$. 
Note that the probability Bob chooses $\task_j$ comes from the cheating strategy; it may not be equal to the probability had he been honest. 
\end{itemize}
   
\paragraph{Where does this idea come from?}  
The idea of the stochastic switch comes directly from stochastic programming. 
Indeed, optimization is an inherent task when studying the security of any cryptographic protocol since we care about an adversary's \emph{optimal} cheating capability. 
In other words, a security analysis often involves maximizing over allowable cheating strategies. 
In this work, we model Alice and Bob's cheating strategies in the same way Kitaev modeled those for coin flipping~\cite{kitaev2002quantum} which involves  semidefinite programming. 

Semidefinite programming is the study of optimizing linear functions over positive semidefinite variables subject to affine constraints. 
Since the study of quantum computing (and hence quantum cryptography) often involves positive semidefinite objects (density operators, POVMs, channels, etc.) this is a very natural setting in which to optimize quantum quantities. 

A semidefinite program (SDP) can be written in a standard form as
\begin{equation}
    \begin{aligned}
        \text{maximize:} & \enspace \inner{C}{X}\\
        \text{subject to:} & \enspace \Phi(X) = B \\
        & \enspace X \succcurlyeq 0, 
    \end{aligned}
\end{equation} 
where $C$ and $B$ are Hermitian matrices, and $\Phi$ is a Hermiticity-preserving linear map. 
An important feature of SDPs is that they can often be solved efficiently and many computational solvers exist~\cite{sturm1999using, toh1999sdpt3, mosek}. 

A stochastic semidefinite program is a variant of an SDP 
where decisions are made during multiple stages based on certain scenarios chosen probabilistically by the adversary 
at each stage. 
(The protocols devised in this work are primarily based on two stages with only a finite number of possible scenarios where each scenario corresponds to a certain protocol.)  
The optimization for a decision-maker within these stochastic protocols is based on the following formulation:   
\begin{equation}
    \begin{aligned}
        \text{maximize:} & \enspace \mathbb{E}_{\omega}[\inner{C_{\omega}}{Y_{\omega}}]\\
        \text{subject to:} & \enspace \Phi_{\omega}(Y_{\omega}) = B_{\omega}, \forall \omega \in \Omega \\
        & \enspace \Xi_{\omega}(Y_{\omega}) = X, \forall \omega \in \Omega \\
        & \enspace Y_{\omega} \succcurlyeq 0, \forall \omega \in \Omega \\
        & \enspace X \succcurlyeq 0, 
    \end{aligned}
\end{equation} 
where $\Omega$ is the set of possible scenarios, $C_{\omega}$ and $B_{\omega}$ are Hermitian for each $\omega$, and $\Phi_{\omega}$ and $\Xi_{\omega}$ are Hermiticity-preserving linear maps for each $\omega$.  

For a finite number of scenarios ($\Omega$ is finite), the above formulation is also an SDP where $X$ is the Stage-I decision made by the decision-maker before the adversary reveals the scenario $\omega$ and $Y_{\omega}$ is the Stage-II decision made when they follow the protocol corresponding to scenario $\omega$. Due to the stochastic nature of scenario selection by the adversary, the decision-maker wishes to select $X$ that maximizes his / her average (expected) probability of success and later takes a decision $Y_{\omega}$ which is optimal for the revealed scenario $\omega$ and chosen $X$. 
The binding constraints of the type $\Xi_{\omega}(Y_{\omega}) = X$ for the Stage-I decision $X$ may restrict decision-maker's success probability compared to the deterministic case where they follow the protocol for scenario $\omega$ with certainty.     
  
\subsection{Our results} 
Our main results can be divided into two categories, breadth and depth. 

\paragraph{Breadth.}
We first examine protocols that provide the ability to switch between different tasks. As discussed previously, here \emph{switch} refers to the functionality where a party can choose to perform one of the tasks (from the allowed set of tasks) after some initial communication with the other party. Here we consider different combinations of the allowed tasks where a task can either be oblivious transfer, bit commitment, \emph{or} weak coin flipping.

We collect all the combinations in the theorem below.

\begin{theorem}[Switching between different protocols, informal]
There exists a simple protocol which performs bit commitment \emph{or} oblivious transfer with 
\begin{equation} 
P_A \approx 0.72855 
\quad \text{ and } \quad 
P_B = 0.75. 
\end{equation}
There exists a simple protocol which performs bit commitment \emph{or} weak coin flipping with 
\begin{equation} 
P_A \approx 0.74381 
\quad \text{ and } \quad 
P_B = 0.75. 
\end{equation} 
There exists a simple protocol which performs weak coin flipping \emph{or} oblivious transfer with 
\begin{equation} 
P_A \approx 0.70440 
\quad \text{ and } \quad 
P_B = 0.75. 
\end{equation} 
There exists a simple protocol which performs bit commitment \emph{or} weak coin flipping \emph{or} oblivious transfer with 
\begin{equation} 
P_A \approx 0.71777 
\quad \text{ and } \quad 
P_B = 0.75. 
\end{equation} 
\end{theorem} 

To give some context, each of the protocols above were achieved by starting with two (or more) protocols, each with Alice and Bob's cheating probabilities equal to $3/4$ then having Bob stochastically switch to thwart Alice's cheating attempts. 
Note that in this case, Bob's cheating probability is not expected to go down since he has the power to select whichever task he wants. 
Thus, the fact that $P_B = 3/4$ still and $P_A < 3/4$  means a strict improvement in the overall security in each case. 

To illustrate the importance of the above theorem, we point out the protocol for oblivious transfer \emph{or} bit commitment. 
In that setting, the two protocols that we switch between are either the best known (as is the case with oblivious transfer~\cite{chailloux2013lowerot}) or the best possible in that setting (as is the case with bit commitment~(\cite{nayak2003bit} based on~\cite{ambainis2001new})).  
At the end, at least one of the two tasks were performed with a strict decrease in Alice's cheating probability (on average we have $0.75$ decreasing to $0.72855$). 
This gives us a strictly better protocol in this sense. 
Moreover, in each of the cases, Alice's average cheating probability decreased below $0.75$, which is better than her standalone cheating probability in each task. 
In summary, our switching idea circumvents a significant barrier on the design of quantum protocols in each case. 

\paragraph{Depth.} 
We next consider switching between two protocols both of which perform the same task. 
In such a case, any stochastic switch of Bob is guaranteed to be a complete protocol for said task. 
The task we consider here is Rabin oblivious transfer. 

As far as we are aware, there are no quantum protocols for Rabin oblivious transfer in the literature. 
Indeed, they are hard to design and even defining what cheating Alice and cheating Bob would want to achieve is a bit tricky. 
However, one can easily define \emph{bad} quantum protocols, in the sense that they are completely broken (and seemingly useless). 
We show that our framework is capable of taking broken protocols and creating some with decent security. 

\begin{theorem}[A quantum protocol for Rabin oblivious transfer, informal] 
    There exists a quantum protocol for Rabin oblivious transfer where Alice can correctly guess whether Bob received the message or learnt $\perp$ (the output where he receives nothing) with probability at most $0.9330$ and Bob can learn Alice's bit with probability at most $0.9691$.
\end{theorem} 

We subsequently extend our framework to develop a quantum protocol for an alternative Rabin oblivious transfer task where dishonest Alice attempts to force Bob to receive nothing.

\begin{theorem}[A quantum protocol for an alternative Rabin oblivious transfer, informal] 
    There exists a quantum protocol where Alice can force Bob to accept $\perp$ (the output where he learns nothing) with probability at most $\cos^2(\pi/8) \approx 0.8535$ and Bob can learn Alice's bit with probability at most $0.875$. 
\end{theorem} 

An interesting aspect of our protocol is that it switches between two protocols where the information is sent only in one direction. 
This allows for an easier analysis and helps to define what a cheating Alice and a cheating Bob would want in this setting. 
We discuss this protocol and Rabin oblivious transfer in detail in Section~\ref{section:depth}. 

\paragraph{Limitations.} 
It is indeed the case that switching does not always yield a more secure protocol. 
We now discuss two examples, one in which switching does not help and one in which it completely breaks the protocol. 

\begin{lemma}[A protocol in which the stochastic switch does not help, informal]
There exists a protocol for XOR oblivious transfer \emph{or} trit commitment in which Bob does the switching but Alice's cheating probability does not decrease. 
\end{lemma} 

We define these two tasks and the two protocols in \cref{appendix:xot:dieroll} followed by the analysis of their stochastic switch. 

\begin{lemma}[A protocol in which the stochastic switch hurts the overall security, informal]
There exists a switching protocol for \emph{strong} coin flipping in which Bob cannot cheat perfectly in each subtask but can cheat perfectly using his stochastic switch. 
\end{lemma} 

We provide the two protocols and the analysis of their stochastic switch in \cref{appendix:bitCommitment:EPR}. 

\subsection{Layout of the paper} 

We start by defining notation and background in \cref{section:background}. 
In \cref{section:baseProtocols} we detail three fundamental two-party cryptographic tasks and in each give a simple quantum protocol along with a full security analysis via semidefinite programming. 
We then consider a stochastic switch between every combination of these fundamental tasks in \cref{section:switchProtocols} and analyze their security using stochastic semidefinite programming. 
In \cref{section:depth} we introduce Rabin oblivious transfer and  use our framework to give the first quantum protocol for this task. 
\cref{appendix:xot:dieroll} and \cref{appendix:bitCommitment:EPR} discuss two instances where the stochastic switch is applicable but does not help in designing protocols with improved security. 
 

\section{Background} \label{section:background}

This section provides a brief background on the mathematical ideas used in this work. We first introduce some notations from linear algebra and matrix analysis followed by a short primer on stochastic semidefinite programming. For a detailed understanding on the relevant subject, we provide a few standard references within the related subsection.  

\subsection{Linear algebra notation and terminology}

We denote the computational (standard) basis for $\mathbb{C}^n$ by $\{\ket{0}, \ket{1}, \ldots \ket{n-1}\}$. We use the Dirac notation from quantum mechanics to represent a vector $\ket{\psi} \in \mathbb{C}^{n}$ and uppercase calligraphic letters (such as $\A, \B$) for denoting complex Euclidean spaces.

For a complex Euclidean space $\mathcal{X}$, the set of operators on $\mathcal{X}$ is given by $\textup{L}_{\mathcal{X}}$ and we define $\mathbbm{1}_{\mathcal{X}}$ as the identity operator acting on $\mathcal{X}$. The set of Hermitian operators on $\mathcal{X}$ is denoted by $\textup{Herm}(\mathcal{X})$ where an operator $X \in \textup{L}_{\mathcal{X}}$ is Hermitian if it is equal to its conjugate-transpose (i.e. $X = X^{\dagger}$). The convex cone of positive semidefinite operators acting on $\mathcal{X}$ is denoted by $\textup{Pos}(\mathcal{X})$ while $\D(\mathcal{X})$ is the set of positive semidefinite operators with unit trace i.e., density operators. The Hilbert-Schmidt inner product between two operators $P,Q \in \textup{Herm}(\mathcal{X})$ is given by
\begin{equation*}
    \inner{P}{Q} = \Tr(P^{\dagger}Q) = \Tr(PQ).
\end{equation*}
The partial trace over $\mathcal{X}$, denoted by $\Tr_{\mathcal{X}}$ is the linear mapping from $\textup{L}_{\mathcal{X} \otimes \mathcal{Y}}$ to $\textup{L}_{\mathcal{Y}}$ such that $\Tr_{\mathcal{X}}(X \otimes Y) = \Tr(X)Y$ for all $X \in \textup{L}_{\mathcal{X}}$ and $Y \in \textup{L}_{\mathcal{Y}}$. 
We also make use of the fact that the measurement of a quantum state $\rho \in  \D(\mathcal{X})$ by a $k$-outcome POVM (positive operator-valued measure) $\{M_0, M_1 \ldots M_{k-1}\}$ reveals outcome $i \in \{0,1,\ldots k-1\}$ with probability $\inner{\rho}{M_i}$ (known as Born's rule).

For a more comprehensive review of quantum information theory, we refer the reader to~\cite{nielsen2010quantum, watrous2018theory}.

\subsection{Stochastic semidefinite programming}
Semidefinite programming (SDP) is a sibling framework of linear programming where we want to optimize a linear function of a positive semidefinite variable constrained to affine subspaces. A popular representation of an SDP is given by:

\begin{equation} \label{sdp:general}
    \begin{aligned}
        \text{maximize:} & \enspace \inner{C}{X}\\
        \text{subject to:} & \enspace \Phi(X) = B \\
        & \enspace X \in \textup{Pos}(\mathcal{X}),
    \end{aligned}
\end{equation}
where $\Phi: \textup{L}(\mathcal{X}) \rightarrow \textup{L}(\mathcal{Y})$ is a Hermiticity-preserving linear map with $C \in \textup{Herm}(\mathcal{X})$ and $B \in \textup{Herm}(\mathcal{Y})$. A semidefinite program can often be solved for an optimal solution using interior point methods~\cite{helmberg1996interior, alizadeh1998primal} efficiently. We refer the reader to~\cite{boyd2004convex} for additional details on semidefinite programming and its theory.  

Stochastic semidefinite programming is a well-studied modelling framework for problems with some degree of stochasticity in the evolution of the underlying environment. We consider a \emph{two-stage} stochastic environment that can be modelled using the following stochastic semidefinite program:
\begin{equation} \label{sdp:stochastic}
    \begin{aligned}
        \text{maximize:} & \enspace \mathbb{E}_{\omega}[\inner{C_{\omega}}{Y_{\omega}}]\\
        \text{subject to:} & \enspace \Phi_{\omega}(Y_{\omega}) = B_{\omega}, \forall \omega \in \Omega \\
        & \enspace \Xi_{\omega}(Y_{\omega}) = X, \forall \omega \in \Omega \\
        & \enspace Y_{\omega} \in \textup{Pos}(\mathcal{Y}_{\omega}), \forall \omega \in \Omega \\
        & \enspace X \in \textup{Pos}(\mathcal{X}).
    \end{aligned}
\end{equation}

Here $X$ is the \emph{here-and-now} (or {Stage-I}) variable whose value is assigned by the decision-maker before the adversary reveals the scenario $\omega$ (which occurs with probability $\mathbb{P}_{\omega}$). $Y_{\omega}$ is the \emph{recourse} (or {Stage-II}) variable assigned by the decision-maker once $\omega$ is revealed. Note that the binding constraints $\Xi_{\omega}(Y_{\omega}) = X$ could restrict the decision-maker to achieve the maximum possible reward (objective value) for scenario $\omega$. The maximum objective value of the program above is bounded above by $\max \limits_{\omega \in \Omega}\{f^*_{\omega}\}$ where $f^*_{\omega}$ is the optimal objective function value for scenario $\omega$ given by the following SDP:
\begin{equation} \label{sdp:scenario}
    \begin{aligned}
        \text{maximize:} & \enspace \inner{C_{\omega}}{Y_{\omega}}\\
        \text{subject to:} & \enspace \Phi_{\omega}(Y_{\omega}) = B_{\omega} \\
        & \enspace Y_{\omega} \in \textup{Pos}(\mathcal{Y}_{\omega}), \\
    \end{aligned}
\end{equation}
i.e., the maximum value in the absence of the binding variable in scenario $\omega$.

If the set of allowed scenarios $\Omega$ is a discrete set, then the formulation~\ref{sdp:stochastic} can be expressed as a regular semidefinite program~\ref{sdp:general}. Often if the number of scenarios (or $|\Omega|$) for a given stochastic SDP is large, then one can exploit the block structure of the SDP to solve it efficiently using the iterative Bender's decomposition~\cite{mehrotra2007decomposition}. As the number of scenarios discussed in this work is relatively small, we resort to solving our stochastic programs using standard SDP solvers. 

\section{Fundamental tasks and simple protocols}\label{section:baseProtocols}

In this section, we discuss some simple protocols for three fundamental cryptographic tasks - bit commitment, coin flipping, and oblivious transfer. 
\subsection{Bit commitment}\label{subsection:bitCommitment}

Bit commitment is a cryptographic task between Alice and Bob who communicate in two phases, a \emph{commit} phase and a \emph{reveal} phase.
\begin{itemize}
    \item In the \emph{commit} phase, Alice communicates with Bob in order to commit to a bit $y \in \{0,1\}$.
    \item In the \emph{reveal} phase, Alice communicates with Bob to reveal her choice bit $y$. We say that Alice successfully reveals if Bob accepts the revealed bit.
\end{itemize}
We define the following notions of security for a valid bit commitment protocol.
\begin{itemize}
    \item \emph{Completeness}: If both Alice and Bob are honest, then Bob always accepts Alice's revealed bit.
    
    \item \emph{Cheating Alice}: A dishonest Alice tries to reveal $\hat{y} \in \{0,1\}$ chosen uniformly at random before the \emph{reveal} phase. The cheating probability of Alice is given by
    	\begin{equation}
        	P_{A}^{BC} = {} \max \frac{1}{2}\Pr[\text{Alice successfully reveals $\hat{y} = 0$}] + \frac{1}{2}\Pr[\text{Alice successfully reveals $\hat{y} = 1$}],
        \end{equation}
        where the maximum is taken over all cheating strategies of Alice. Note that $P_{A}^{BC} \geq 1/2$ since she can always make an honest commit to $y$ in the \emph{commit} phase which is successfully revealed with probability $1$ when $\hat{y} = y$ (which occurs with probability $1/2$).     
	  
    \item \emph{Cheating Bob}: A dishonest Bob tries to learn the commit bit before the \emph{reveal} phase. The cheating probability of Bob is given by
    \begin{equation}
        P_{B}^{BC} = {} \max \Pr[\text{Bob learns $y$ before the reveal phase}],
    \end{equation}
    where the maximum is taken over all cheating strategies of Bob. Note that $P_B^{BC} \geq 1/2$ since Bob can trivially guess $y$ with probability $1/2$ by selecting one of the two possible values uniformly at random.
\end{itemize}

In some previous works, \cite{lo1997quantum}, \cite{mayers1997unconditionally} showed that a perfectly secure bit commitment with quantum communication is impossible while more recently, \cite{sikora2018bitcommitment} provided a simple proof of this impossibility for generalized probabilistic theories. In another work, \cite{chailloux2011bc} extended the optimal strong coin flipping protocol \cite{chailloux2009scf} to come up with an optimal protocol with a value for $0.739$ on $\max\{P_A^{BC}, P_B^{BC}\}$. However, these protocols are quite complicated and require an infinite number of messages (in the limit). 

We now present a simple quantum bit commitment protocol.
\begin{protocol}{Quantum protocol for bit commitment \cite{kerenidis2004weak}.}\label{protocol:bitCommitment}
    \begin{itemize}
        \underline{Stage-I}
        \item (Commitment) \textbf{Alice chooses a bit $y \in \{0,1\}$ uniformly at random and creates the two-qutrit state 
        \begin{equation*}
            \ket{\phi_y} = \frac{1}{\sqrt{2}}\ket{yy} + \frac{1}{\sqrt{2}}\ket{22} \in \A \otimes \B.
        \end{equation*}
        \item Alice sends the qutrit $\mathcal{B}$ to Bob.}\par
        \underline{Stage-II}
        \item (Revelation) Alice reveals $y$ to Bob and sends him the subsystem $\A$.
        \item (Verification) Bob measures the combined state in $\A \otimes \B$ to accept or reject with the POVM:
        \begin{equation*}
            \{\Pi_{accept} \coloneqq \ketbra{\phi_y}{\phi_y}_{\A \otimes \B}, \Pi_{reject} \coloneqq \mathbbm{1}_{\A \otimes \B} - \Pi_{accept} \}.
        \end{equation*}
        If Bob accepts, Alice successfully reveals $y$.
    \end{itemize}
\end{protocol}

\subsubsection{Cheating Alice}
In order for Alice to successfully reveal bit $\hat{y} \in \{0,1\}$ in \cref{protocol:bitCommitment}, she would send $\hat{y}$ and the qutrit $\A$ such that Bob hopefully accepts the combined state of the two qutrits (in $\A \otimes \B$) with high probability. It is instructive to note that $\hat{y}$ is sampled uniformly at random after Alice has already sent $\B$ during the \emph{commit} phase.

The success probability of Alice can be evaluated by considering the two equally likely scenarios for the observed $\hat{y}$, for each of which Alice wishes a successful verification by Bob's measurement. If $\sigma_{\hat{y}}^{BC}$ is the final state in $\D(\A \otimes \B)$ with Bob after the \emph{reveal} phase, then the probability of successful verification (accept) is given by $\inner{\sigma_{\hat{y}}^{BC}}{\ketbra{\phi_{\hat{y}}}{\phi_{\hat{y}}}}$. As the qutrit $\B$ was already communicated to Bob in the \emph{commit} phase independent of $\hat{y}$, the reduced state $\B$ in the final state $\sigma_{\hat{y}}^{BC}$ is independent of $\hat{y} \in \{0,1\}$ implying $\Tr_{\A}(\sigma_{0}^{BC}) = \Tr_{\A}(\sigma_{1}^{BC})$. The optimal cheating probability can thus be calculated by maximizing the probability $\inner{\sigma_{\hat{y}}^{BC}}{\ketbra{\phi_{\hat{y}}}{\phi_{\hat{y}}}}$ averaged over both values of $\hat{y}$.

\textbf{SDP for cheating Alice.} Alice's optimal cheating probability $P_A^{BC}$ is given by the optimal objective function value of the following SDP:
\begin{equation}\label{sdp:bitCommitment:Alice}
    \begin{aligned}
        \text{maximize:} & \enspace \frac{1}{2}\inner{\sigma_0^{BC}}{\kb{\phi_0}} + \frac{1}{2}\inner{\sigma_1^{BC}}{\kb{\phi_1}}\\
        \text{subject to:} & \enspace \Tr_{\A}(\sigma_0^{BC}) = \Tr_{\A}(\sigma_1^{BC}) = \sigma^{BC} \\
        & \enspace \sigma_0^{BC}, \sigma_1^{BC} \in \D(\A \otimes \B) \\
        & \enspace \sigma^{BC} \in \D(\B).
    \end{aligned}
\end{equation}

Numerically solving the above SDP, we find that Alice can cheat with an optimal probability of $P_A^{BC} = 3/4$ and this is achieved when $\sigma^{BC} = 
\begin{bmatrix}
    1/6 & 0 & 0 \\
    0 & 1/6 & 0 \\
    0 & 0 & 2/3
\end{bmatrix}$.

\subsubsection{Cheating Bob}
Typically, a security analysis for dishonest Bob involves him performing a state discrimination on the two possible states that could be sent during the \emph{commit} phase.
A slightly different way to quantify the success probability of Bob in identifying Alice's commit bit is to introduce a verification process on Alice's side where Bob sends back his guess for the state chosen by Alice and is subsequently verified by her for its correctness. Note that this is just a hypothetical phase introduced to capture Bob's cheating probability; it is not part of the protocol. 

Mathematically, Alice prepares a pure state $\ket{\psi} = \sum \limits_{y \in \{0,1\}}\frac{1}{\sqrt{2}}\ket{y}_{\mathcal{Y}}\ket{\phi_y}_{\A \otimes \B}$ at the beginning of the protocol and sends the qutrit $\B$ to Bob who subsequently returns a guess $g$ in $\mathcal{G}$. Alice checks if $g = y$ using the POVM: $$\{ Q_{accept} \coloneqq \sum \limits_{g \in \{0,1\}}\sum \limits_{y \in \{0,1\}} \delta_{y,g}\kb{y}_{\mathcal{Y}} \otimes \mathbbm{1}_{\A} \otimes \kb{g}_{\mathcal{G}}, Q_{reject} \coloneqq \mathbbm{1}_{\mathcal{Y} \otimes \A \otimes \mathcal{G}} - Q_{accept} \}.$$
The optimal probability of correctly guessing $y$ is obtained by maximizing $\inner{\tau^{BC}}{Q_{accept}}$ where $\tau^{BC}$ is the state with Alice in $\D(\mathcal{Y} \otimes \A \otimes \mathcal{G})$ after receiving the guess $g$ in $\mathcal{G}$ from Bob. 

\textbf{SDP for cheating Bob.} The optimal cheating probability for Bob is given by the optimal objective function value of the following SDP:
\begin{equation}\label{sdp:bitCommitment:Bob}
    \begin{aligned}
        \text{maximize:} & \enspace \inner{\tau^{BC}}{Q_{accept}} \\
        \text{subject to:} & \enspace \Tr_{\mathcal{G}}(\tau^{BC}) = \Tr_{\B}(\kb{\psi}) \\
        & \enspace \tau^{BC} \in \D(\mathcal{Y} \otimes \A \otimes \mathcal{G})
    \end{aligned}
\end{equation}
where recall, $\ket{\psi} = \sum \limits_{y \in \{0,1\}}\frac{1}{\sqrt{2}}\ket{y}_{\mathcal{Y}}\ket{\phi_y}_{\A \otimes \B}$ and $$Q_{accept} = \sum \limits_{g \in \{0,1\}} \sum \limits_{y \in \{0,1\}} \delta_{y,g} \kb{y}_{\Y} \otimes \mathbbm{1}_{\A} \otimes \kb{g}_{\mathcal{G}}.$$
    
Numerically solving the above SDP, we find that Bob can cheat with an optimal probability of $P_B^{BC} = 3/4$.

\subsection{Weak coin flipping}\label{subsection:coinFlip:EPR}

Coin flipping is the cryptographic task between two parties (Alice and Bob) where they communicate to agree on a common binary outcome ($0$ or $1$). A weak version of this task is a variant where Alice \emph{wins} if the common outcome is $1$ while Bob \emph{wins} if the outcome is $0$.

We consider the following notions of security for a given weak coin flipping protocol.
\begin{itemize}
    \item \emph{Completeness:} If both Alice and Bob are honest, then neither party aborts and the shared outcome is generated uniformly at random. 
    \item \emph{Cheating Alice:} If Bob is honest, then Alice's cheating probability is defined as
    \begin{equation*}
        P_{A}^{WCF} = \max \Pr[\text{Alice successfully forces the outcome $1$}],
    \end{equation*}
    where the maximum is taken over all possible cheating strategies of Alice. Note that here $P_A^{WCF} \geq 1/2$ since Alice can simply choose to follow the protocol honestly to observe the outcome $1$ uniformly at random.
    \item \emph{Cheating Bob:} If Alice is honest, then Bob's cheating probability is defined as
    \begin{equation*}
        P_{B}^{WCF} = \max \Pr[\text{Bob successfully forces the outcome $0$}],
    \end{equation*}
    where the maximum is taken over all possible cheating strategies of Bob. As before, here $P_B^{WCF} \geq 1/2$ since Bob can simply choose to follow the protocol honestly to observe the outcome $0$ uniformly at random. 
\end{itemize}
In a previous work, \cite{mochon2007quantum} showed that the lower bound of $1/2$ on $\max\{P_A^{WCF}, P_B^{WCF}\}$ is in fact tight up to any precision $\epsilon$. More recently, \cite{miller2020impossibility} established the impossibility of efficient coin flipping by developing lower bounds of $\exp(\Omega(1/\sqrt{\epsilon}))$ on the communication complexity.

\newpage
We now present a simple weak coin flipping protocol.
\begin{protocol}{Quantum protocol for coin flipping based on EPR \cite{colbeck2007entanglement}.}\label{protocol:coinFlip:EPR}
    \begin{itemize}
        \underline{Stage-I}
        \item \textbf{Alice chooses a bit $y \in \{0,1\}$ uniformly at random and creates the two-qutrit state 
        \begin{equation*}
            \ket{\phi_y} = \frac{1}{\sqrt{2}}\ket{yy} + \frac{1}{\sqrt{2}}\ket{22} \in \A_0 \otimes \B_0.
        \end{equation*}
        \item Alice sends the qutrit $\mathcal{B}_0$ to Bob}.\par
        \underline{Stage-II}
        \item Alice creates another two-qutrit state
        \begin{equation*}
            \ket{\phi_y} = \frac{1}{\sqrt{2}}\ket{yy} + \frac{1}{\sqrt{2}}\ket{22} \in \A_1 \otimes \B_1
        \end{equation*}
        and sends the qutrit $\B_1$ to Bob.
        \item Bob chooses an independent bit $z \in \{0,1\}$ uniformly at random and sends it to Alice.
        \item Alice reveals $y$ to Bob and sends him the qutrit $\A_z$.
        \item Bob measures the combined state $\A_z \otimes \B_z$ to accept or reject with the POVM:
        \begin{equation*}
            \{\Pi_{accept} \coloneqq \ketbra{\phi_y}{\phi_y}_{\A_{z} \otimes \B_{z}}, \Pi_{reject} \coloneqq \mathbbm{1}_{\A_z \otimes \B_z} - \Pi_{accept} \}.
        \end{equation*}
        \item If Bob accepts, Alice and Bob measures their respective subsystems with the POVM:
        \begin{equation*}
            \{\Pi_{0} \coloneqq \kb{0} + \kb{1}, \Pi_{1} \coloneqq \kb{2} \}
        \end{equation*}
        to get the outcome of the protocol.
    \end{itemize}
\end{protocol}

\subsubsection{Cheating Alice}
The idea here is that dishonest Alice attempting to force the outcome $1$ in \cref{protocol:coinFlip:EPR} would want Bob to accept the combined state in $\A_z \otimes \B_z$ and obtain the measurement outcome $1$ on the subsystem $\B_{\overline{z}}$.
As Bob measures just the qutrit $\B_{\overline{z}}$ to obtain his outcome, we can assume that Alice sends both $\A_z \otimes \A_{\overline{z}}$ (instead of just $\A_z$) to Bob.
Note that this does not affect the success probability of Alice and greatly simplifies her overall security analysis. 
The combined success probability can be evaluated by the inner product $\inner{\sigma_z^{WCF}}{P_z}$ where $\sigma_z^{WCF}$ is the state with Bob after receiving $y$ and $\A_z \otimes \B_z$ where $$P_z = \sum \limits_{y \in \{0,1\}} \kb{y}_{\mathcal{Y}} \otimes \kb{\phi_y}_{\A_z \otimes \B_z} \otimes \mathbbm{1}_{\A_{\overline{z}}} \otimes \kb{2}_{\B_{\overline{z}}}$$ is the joint measurement that evaluates the combined probability of Bob accepting the state $\ket{\phi_y}$ in $\A_z \otimes \B_z$ and obtaining the outcome $1$ on $\B_{\overline{z}}$ for both choices of $y$. 
As the qutrits $\B_0$ and $\B_1$ were sent to Bob at the beginning of the protocol before $y$ and $\A_0,\A_1$, the reduced state in $\B_0 \otimes \B_1$ is independent of them implying $\Tr_{\mathcal{Y} \A_0 \A_1}(\sigma_0^{WCF}) = \Tr_{\mathcal{Y} \A_0 \A_1}(\sigma_1^{WCF})$. The optimal cheating probability is thus calculated by maximizing the probability $\inner{\sigma_z^{WCF}}{P_z}$ averaged over both choices of $z$.

\textbf{SDP for cheating Alice.} The optimal cheating probability for Alice attempting to force the outcome $1$ is given by the optimal objective function value of the following SDP:

\begin{equation}\label{sdp:coinFlip:EPR:Alice}
    \begin{aligned}
        \text{maximize:} & \enspace \frac{1}{2}\inner{\sigma_0^{WCF}}{P_0} + \frac{1}{2}\inner{\sigma_1^{WCF}}{P_1} \\
        \text{subject to:} & \enspace \Tr_{\mathcal{Y} \A_0 \A_1}(\sigma_0^{WCF}) = \Tr_{\mathcal{Y} \A_0 \A_1}(\sigma_1^{WCF}) = \sigma^{WCF}_{B}\\
        & \enspace \Tr_{\B_1}(\sigma^{WCF}_{B}) = \sigma^{WCF} \\
        & \enspace \sigma_0^{WCF}, \sigma_1^{WCF} \in \D(\mathcal{Y} \otimes \A_0 \otimes \B_0 \otimes \A_1 \otimes \B_1) \\
        & \enspace \sigma^{WCF}_{B} \in \D(\B_0 \otimes \B_1) \\
        & \enspace \sigma^{WCF} \in \D(\B_0)
    \end{aligned}
\end{equation}
where,
\begin{equation*}
    \begin{aligned}
        P_0 &= \sum_{y \in \{0,1\}} \kb{y}_{\mathcal{Y}} \otimes \kb{\phi_y}_{\A_0 \otimes \B_0} \otimes \mathbbm{1}_{\A_1} \otimes \kb{2}_{\B_1} \\
        P_1 &= \sum_{y \in \{0,1\}} \kb{y}_{\mathcal{Y}} \otimes \mathbbm{1}_{\A_0} \otimes \kb{2}_{\B_0} \otimes \kb{\phi_y}_{\A_1 \otimes \B_1}.
    \end{aligned}
\end{equation*}

Numerically solving the above SDP, we find that Alice can cheat with an optimal probability of $P_A^{WCF} = 3/4$ and this is achieved when $\sigma^{WCF} = 
\begin{bmatrix}
    1/12 & 0 & 0 \\
    0 & 1/12 & 0 \\
    0 & 0 & 5/6 
\end{bmatrix}
$.

\subsubsection{Cheating Bob}

To analyze Bob's cheating, consider the state with Alice $\ket{\psi} = \sum \limits_{y \in \{0,1\}}\frac{1}{\sqrt{2}}\ket{y}_{\mathcal{Y}}\ket{\phi_y}_{\A_0 \otimes \B_0}\ket{\phi_y}_{\A_1 \otimes \B_1}$ in $\D(\mathcal{Y} \otimes \A_0 \otimes \B_0 \otimes \A_1 \otimes \B_1)$ at the beginning of the protocol. After Alice sends the qutrits $\B_0$ and $\B_1$ to Bob, she receives $z$ in return and is then holding the state $\tau^{WCF} \in \D(\mathcal{Y} \otimes \A_0 \otimes \A_1 \otimes \mathcal{Z})$. Note that Bob could use the qutrits $\B_0$ and $\B_1$ to obtain a preferable choice for $z$ before sending it to Alice. Since the qutrit $\A_z$ sent to Bob in the final message of the protocol does not affect Alice's outcome, one can safely disregard this message just to come up with a simpler formulation for cheating Bob. The reduced state in $\D(\mathcal{Y} \otimes \A_0 \otimes \A_1)$ remains unchanged from the state with Alice at the beginning of the protocol implying $\Tr_{\mathcal{Z}}(\tau^{WCF}) = \Tr_{\B_0 \B_1}(\kb{\psi})$. The probability of Alice observing the outcome $0$ is given by $\inner{\tau^{WCF}}{Q}$ where $$Q = \sum \limits_{z}\mathbbm{1}_{\mathcal{Y}} \otimes \mathbbm{1}_{\A_z} \otimes (\kb{0} + \kb{1})_{\A_{\overline{z}}} \otimes \kb{z}_{\mathcal{Z}}$$ is the combined measurement that evaluates the probability of Alice observing $0$ on $\A_{\overline{z}}$ when Bob sends back $z$. The optimal cheating probability for Bob can thus be calculated by maximizing $\inner{\tau^{WCF}}{Q}$.

\textbf{SDP for cheating Bob.} The optimal cheating probability for Bob to force the outcome $0$ is given by the optimal objective function value of the following SDP:
\begin{equation}\label{sdp:coinFlip:EPR:Bob}
    \begin{aligned}
        \text{maximize:} & \enspace \inner{\tau^{WCF}}{Q} \\
        \text{subject to:} & \enspace \Tr_{\mathcal{Z}}(\tau^{WCF}) = \Tr_{\B_0 \B_1}(\kb{\psi}) \\
        & \enspace \tau^{WCF} \in \D(\mathcal{Y} \otimes \A_0 \otimes \A_1 \otimes \mathcal{Z})
    \end{aligned}
\end{equation}
where recall, $$Q = \sum \limits_{z \in \{0,1\}}\mathbbm{1}_{\mathcal{Y}} \otimes \mathbbm{1}_{\A_z} \otimes (\kb{0} + \kb{1})_{\A_{\overline{z}}} \otimes \kb{z}_{\mathcal{Z}}.$$

Numerically solving the above SDP, we find that Bob can cheat with an optimal probability of $P_B^{WCF} = 3/4$.

\subsection{Oblivious transfer}\label{subsection:obliviousTransfer:1outOf2}

Oblivious transfer (OT) is the fundamental cryptographic task between two parties, Alice and Bob, where Alice wants to learn one of two pieces of information while keeping Bob oblivious to what she learned. We define the task of $1$-out-of-$2$ OT as follows.
\begin{itemize}
    \item Bob selects $(x_0,x_1) \in \{0,1\}^2$ uniformly at random.
    \item Alice selects $y \in \{0,1\}$ uniformly at random.
    \item Alice learns $x_y$.
\end{itemize}

We define the following notions of security for $1$-out-of-$2$ OT protocols.
\begin{itemize}
    \item \emph{Completeness:} A $1$-out-of-$2$ OT protocol is said to be \emph{complete} if Alice learns the selected bit $x_y$ unambiguously. 
     
    \item \emph{Cheating Alice:} A dishonest Alice attempts to learn both $x_0$ and $x_1$. The cheating probability of Alice is given by
    \begin{equation*}
        P_A^{OT} = \max \Pr[\text{Alice correctly guesses both $x_0$ and $x_1$}],
    \end{equation*}
    where the maximum is taken over all cheating strategies of Alice. Note that $P_A^{OT} \geq 1/2$ since she can simply follow the protocol to learn $x_y$ perfectly (completeness) and can correctly guess $x_{\overline{y}}$ by making a guess uniformly at random.

    \item \emph{Cheating Bob:} A dishonest Bob attempts to learn the bit $y$. The cheating probability of Bob is given by
    \begin{equation*}
        P_B^{OT} = \max \Pr[\text{Bob correctly guesses $y$}],
    \end{equation*}
    where the maximum is taken over all cheating strategies of Bob. Note that $P_B^{OT} \geq 1/2$ since he can correctly guess Alice's bit $y$ by guessing one of the two values uniformly at random.
\end{itemize}  

It was demonstrated in \cite{chailloux2013lowerot} a lower bound of $0.5852$ on $\max \{P_A^{OT}, P_B^{OT}\}$ which was subsequently improved to $2/3$ if we consider the setting of semi-honest oblivious transfer \cite{chailloux2016semihonest}. More recently, \cite{amiri2021imperfect} developed tighter lower bounds for the class of OT protocols where the states outputted by Alice, when both parties are honest, are pure and symmetric. The optimality of the existing lower bound of $2/3$ for the general class of OT protocols remains an open question.    

We now reproduce a simple $1$-out-of-$2$ OT protocol.
\begin{protocol}{Quantum protocol for $1$-out-of-$2$ oblivious transfer \cite{chailloux2013lowerot}.}\label{protocol:obliviousTransfer:1outOf2}
    \begin{itemize}
        \underline{Stage-I}
        \item \textbf{Alice chooses a bit $y \in \{0,1\}$ uniformly at random and creates the two-qutrit state 
        \begin{equation*}
            \ket{\phi_y} = \frac{1}{\sqrt{2}}\ket{yy} + \frac{1}{\sqrt{2}}\ket{22} \in \A \otimes \B.
        \end{equation*}
        \item Alice sends the qutrit $\mathcal{B}$ to Bob.}\par
        \underline{Stage-II}
        \item Bob randomly selects $\{x_0, x_1\} \in \{0,1\}^2$ and applies the unitary 
        \begin{equation*}
            U_{x_0x_1} = 
            \begin{bmatrix}
                (-1)^{x_0} & 0 & 0 \\
                0 & (-1)^{x_1} & 0 \\
                0 & 0 & 1
            \end{bmatrix}
        \end{equation*}
        to the received qutrit. Afterwards, Bob returns the qutrit to Alice.
        \item Alice determines the value of $x_y$ using the two-outcome measurement:
        \begin{equation*}
            \{\Pi_{0} \coloneqq \kb{\phi_y}_{\A \otimes \B}, \Pi_{1} \coloneqq \mathbbm{1}_{\A \otimes \B} - \kb{\phi_y}  \}.
        \end{equation*}
    \end{itemize}
\end{protocol}

\subsubsection{Cheating Alice}
A dishonest Alice in \cref{protocol:obliviousTransfer:1outOf2} could attempt to learn both bits simultaneously by creating a state that will eventually encode both bits ($x_0,x_1$). In doing so, Alice might fail to learn her choice bit $x_y$ with certainty but the strategy could improve her overall chances of learning both bits.

To quantify the extent of Alice's cheating, consider the state $$\ket{\psi} = \frac{1}{2}\sum \limits_{(x_0,x_1) \in \{0,1\}^2}\ket{x_0x_1}_{\mathcal{X}_0 \otimes \mathcal{X}_1}$$ held by Bob at the beginning of the protocol. On receiving the qutrit $\B$ from Alice, Bob now holds the joint state $\kb{\psi} \otimes \sigma_{B}^{OT}$ in $\D(\mathcal{X}_0 \otimes \mathcal{X}_1 \otimes \B)$ and applies the unitary $$U_2 = \sum \limits_{x_0,x_1 \in \{0,1\}^2} \kb{x_0x_1} \otimes U_{x_0x_1}$$ (controlled on Bob's choice bits $(x_0,x_1)$) on this joint state.
Once Bob sends back the qutrit $\B$, we extend the strategy similar to the one used previously in \ref{sdp:bitCommitment:Bob} to introduce a hypothetical message where Bob receives the guess $(g_0,g_1)$ for the chosen bits $(x_0,x_1)$ from Alice and holds the state 
$\sigma^{OT}$ in $\D(\mathcal{X}_0 \otimes \mathcal{X}_1 \otimes \mathcal{G}_0 \otimes \mathcal{G}_1)$. Bob measures $\sigma^{OT}$ to check if $(g_0,g_1) = (x_0,x_1)$ using the POVM:   
\begin{equation*}
    \begin{aligned}
        \big\{P_{accept} &\coloneqq \sum \limits_{(g_0,g_1)} \sum \limits_{(x_0,x_1)} \delta_{x_0,g_0} \delta_{x_1,g_1} \kb{x_0x_1}_{\mathcal{X}_0 \otimes \mathcal{X}_1} \otimes \kb{g_0g_1}_{\mathcal{G}_0 \otimes \mathcal{G}_1}, \\ 
        P_{reject} &\coloneqq \mathbbm{1}_{\mathcal{X}_0 \otimes \mathcal{X}_1 \otimes \mathcal{G}_0 \otimes \mathcal{G}_1} - P_{accept}\big\},
    \end{aligned} 
\end{equation*}
where the probability of correctly guessing $(x_0,x_1)$ is given by $\inner{\sigma^{OT}}{P_{accept}}$. As the states $\mathcal{X}_0$ and  $\mathcal{X}_1$ with Bob remain unchanged, we have the constraint $$\Tr_{\mathcal{G}_0\mathcal{G}_1}(\sigma^{OT}) = \Tr_{\B}\Big(U_{2}(\kb{\psi} \otimes \sigma_{B}^{OT})U_{2}^{\dagger}\Big)$$ when maximizing $\inner{\sigma^{OT}}{P_{accept}}$.

\textbf{SDP for cheating Alice.} The optimal cheating probability for Alice to correctly guess $(x_0,x_1)$ is given by the optimal objective function value of the following SDP:
\begin{equation}\label{sdp:obliviousTransfer:1outOf2:Alice}
    \begin{aligned}
        \text{maximize:} & \enspace \inner{\sigma^{OT}}{P_{accept}}  \\
        \text{subject to:} & \enspace \Tr_{\mathcal{G}_0\mathcal{G}_1}(\sigma^{OT}) = \Tr_{\B}\Big(U_{2}(\kb{\psi} \otimes \sigma_{B}^{OT})U_{2}^{\dagger}\Big) \\
        & \enspace \sigma^{OT} \in \D(\mathcal{X}_0 \otimes \mathcal{X}_1 \otimes \mathcal{G}_0 \otimes \mathcal{G}_1) \\
        & \enspace \sigma_{B}^{OT} \in \D(\B)
    \end{aligned}
\end{equation}
where recall,
\begin{equation*}
    \begin{aligned}
        \ket{\psi} &= \frac{1}{2}\sum_{x_0,x_1}\ket{x_0x_1}_{\mathcal{X}_0 \otimes \mathcal{X}_1} \\
        U_2 &= \sum_{x_0,x_1} \kb{x_0x_1}_{\mathcal{X}_0 \otimes \mathcal{X}_1} \otimes U_{x_0x_1} \\
        P_{accept} &= \sum \limits_{(g_0,g_1) \in \{0,1\}^2} \sum \limits_{(x_0,x_1) \in \{0,1\}^2} \delta_{x_0,g_0} \delta_{x_1,g_1} \kb{x_0x_1}_{\mathcal{X}_0 \otimes \mathcal{X}_1} \otimes \kb{g_0g_1}_{\mathcal{G}_0 \otimes \mathcal{G}_1}.
    \end{aligned}
\end{equation*}
Numerically solving the above SDP, we find that Alice can cheat with an optimal probability of $P_A^{OT} = 3/4$ and this is achieved for $\sigma_{B}^{OT} = 
\begin{bmatrix}
    1/3 & 0 & 0 \\
    0 & 1/3 & 0 \\
    0 & 0 & 1/3
\end{bmatrix}$.

\subsubsection{Cheating Bob}
A dishonest Bob on receiving the qutrit $\B$, would devise a two-outcome measurement that maximizes his chances of correctly guessing Alice's choice bit $y$.
The corresponding SDP formulation and the optimal strategy for Bob would be the exact same as discussed in \cref{subsection:bitCommitment} and thus $P_B^{OT}$ evaluates to $3/4$.

\section{Stochastic switches between different protocols}\label{section:switchProtocols}

In this section, we depict various stochastic switches with the protocols discussed in \cref{section:baseProtocols} and analyze their security with the underlying notions described in \cref{subsection:intro:stochasticSwitching}.

\subsection{Stochastic switch between bit commitment and oblivious transfer}

We present the switch protocol below.
\begin{protocol}{Quantum BC-OT stochastic switch protocol.}\label{protocol:switch:bitCommitment:obliviousTransfer:1outOf2}
    \begin{itemize}
        \underline{Stage-I}
        \item \textbf{Alice chooses a bit $y \in \{0,1\}$ uniformly at random and creates the two-qutrit state 
        \begin{equation*}
            \ket{\phi_y} = \frac{1}{\sqrt{2}}\ket{yy} + \frac{1}{\sqrt{2}}\ket{22} \in \A \otimes \B.
        \end{equation*}
        \item Alice sends the qutrit $\mathcal{B}$ to Bob.}\par \newpage
        \underline{Stage-II}
        \item Bob chooses $c \in \{0,1\}$ uniformly at random and sends it to Alice.
        \item If $c = 0$, Alice and Bob perform bit commitment as per \cref{protocol:bitCommitment}, otherwise, they perform $1$-out-of-$2$ oblivious transfer as per \cref{protocol:obliviousTransfer:1outOf2}. 
    \end{itemize}
\end{protocol}

\subsubsection{Cheating Alice}
Dishonest Alice in \cref{protocol:switch:bitCommitment:obliviousTransfer:1outOf2} would like to maximize her average chances of successfully revealing $\hat{y}$ or correctly guessing $(x_0,x_1)$ when Bob decides on either bit commitment or oblivious transfer (with equal probability) respectively.
The constraints formulated in the SDPs for these two tasks in \ref{sdp:bitCommitment:Alice} and \ref{sdp:obliviousTransfer:1outOf2:Alice} should be jointly satisfied as Bob can randomly select one of the two tasks. We introduce an additional constraint in the form of $\sigma^{BC} = \sigma_{B}^{OT}$ to convey the fact that the message sent by Alice in the beginning of the protocol remains independent of Bob's selection of $c$ in Stage-II of the switch protocol. We finally maximize the average success of Alice for the two tasks under the above constraints.

\textbf{SDP for cheating Alice.} The maximum cheating probability for Alice is given by the optimal objective function value of the following SDP:
\begin{flalign*}
    \text{maximize: } \frac{1}{2}\bigg(\frac{1}{2}\inner{\sigma_0^{BC}}{\kb{\phi_0}} + \frac{1}{2}\inner{\sigma_1^{BC}}{\kb{\phi_1}}\bigg) + \frac{1}{2}\inner{\sigma^{OT}}{P_{accept}} &&
\end{flalign*}
subject to: \\
\begin{minipage}{0.50\linewidth}
    \begin{equation*}
        \begin{aligned}
            & \enspace \Tr_{\A}(\sigma_0^{BC}) = \Tr_{\A}(\sigma_1^{BC}) = \sigma^{BC} \\
            & \enspace \sigma_0^{BC}, \sigma_1^{BC} \in \D(\A \otimes \B) \\
            & \enspace \sigma^{BC} \in \D(\B)
        \end{aligned}
    \end{equation*}
\end{minipage}%
\begin{minipage}{0.50\linewidth}
    \begin{equation*}
        \begin{aligned}
            & \Tr_{\mathcal{G}_0\mathcal{G}_1}(\sigma^{OT}) = \Tr_{\B}\Big(U_{2}(\kb{\psi} \otimes \sigma_{B}^{OT})U_{2}^{\dagger}\Big) \\
            & \sigma^{OT} \in \D(\mathcal{X}_0 \otimes \mathcal{X}_1 \otimes \mathcal{G}_0 \otimes \mathcal{G}_1) \\
            & \sigma_{B}^{OT} \in \D(\B)
        \end{aligned}
    \end{equation*}
\end{minipage}
\begin{equation*}
    \boxed{\sigma^{BC} = \sigma_{B}^{OT}},
\end{equation*}
where $\ket{\phi_0}$ and $\ket{\phi_1}$ are defined in \cref{sdp:bitCommitment:Alice} and $P_{accept}$, $U_2$ and $\ket{\psi}$ are as defined in \cref{sdp:obliviousTransfer:1outOf2:Alice}. \\

Numerically solving the above SDP, we find that Alice can cheat with an optimal probability of $P_A = 0.728557$ and this is achieved when $\sigma^{BC} = \sigma_{B}^{OT} = 
\begin{bmatrix}
    0.25 & 0 & 0 \\
    0 & 0.25 & 0 \\
    0 & 0 & 0.5 \\
\end{bmatrix}$.

We remark that $P_A < P_A^{BC}$ and $P_A < P_A^{OT}$ implying an improvement in the security of the switch protocol compared to just \cref{protocol:bitCommitment} or \cref{protocol:obliviousTransfer:1outOf2}.

\subsubsection{Cheating Bob}
Dishonest Bob in \cref{protocol:switch:bitCommitment:obliviousTransfer:1outOf2} would like to correctly guess $y$ for both the tasks of bit commitment and oblivious transfer. Mathematically, Alice holds the state $\tau$ in $\D(\mathcal{C} \otimes \mathcal{Y} \otimes \A \otimes \mathcal{G})$ after sending the qutrit $\B$ and receiving the choice state in $c$ for the selected protocol along with the guess $g$ for $y$. As the qutrit $\A$ and the state for $y \in \mathcal{Y}$ remains unchanged from the state with Alice $\ket{\psi} = \sum \limits_{y} \frac{1}{\sqrt{2}}\ket{y}_{\mathcal{Y}} \ket{\phi_y}_{\A \otimes \B}$ at the beginning of the protocol, we introduce a representative constraint of the form $\Tr_{\mathcal{C} \mathcal{G}}(\tau) = \Tr_{\B}(\kb{\psi})$. We finally maximize the probability of Bob correctly guessing $y$ using the appropriate measurements for the two possible choices of $c$. Note that dishonest Bob can set $c$ via postselection on a measurement on the qutrit $\B$.

\textbf{SDP for cheating Bob.} The maximum cheating probability for Bob is given by the optimal objective function value of the following SDP:
\begin{equation}\label{sdp:switch:bitCommitment:obliviousTransfer:Bob}
    \begin{aligned}
        \text{maximize:} & \enspace \inner{\tau}{Q^{BC}} + \inner{\tau}{Q^{OT}}  \\
        \text{subject to:} & \enspace \Tr_{\mathcal{C} \mathcal{G}}(\tau) = \Tr_{\B}(\kb{\psi}) \\
        & \enspace \tau \in \D(\mathcal{C} \otimes \mathcal{Y} \otimes \A \otimes \mathcal{G})
    \end{aligned}
\end{equation}
where, 
\begin{equation*}
    \begin{aligned}
        \ket{\psi} &= \sum_{y \in \{0,1\}} \frac{1}{\sqrt{2}}\ket{y}_{\mathcal{Y}} \ket{\phi_y}_{\A \otimes \B} \\
        Q^{BC} &= \kb{0}_{\mathcal{C}} \otimes \Big(\sum_{(y,g) \in \{0,1\}^2} \delta_{y,g} \kb{y}_{\mathcal{Y}} \otimes \mathbbm{1}_{\A} \otimes \kb{g}_{\mathcal{G}}\Big) \\
        Q^{OT} &= \kb{1}_{\mathcal{C}} \otimes \Big(\sum_{(y',g') \in \{0,1\}^2} \delta_{y,g }\kb{y'}_{\mathcal{Y}} \otimes \mathbbm{1}_{\A} \otimes \kb{g'}_{\mathcal{G}}\Big).
    \end{aligned}
\end{equation*}
Numerically solving the above SDP, we find that Bob can cheat with an optimal probability of $P_B = 3/4$. Note that this is not surprising since Bob can cheat with a maximum probability of $3/4$ in the constituent protocols as both the protocols require him to correctly guess Alice's state once he receives qutrit $\B$.

\subsection{Stochastic switch between bit commitment and weak coin flipping}

We present the switch protocol below.
\begin{protocol}{Quantum BC-WCF stochastic switch protocol.}\label{protocol:switch:bitCommitment:coinFlip:EPR}
    \begin{itemize}
        \underline{Stage-I}
        \item \textbf{Alice chooses a bit $y \in \{0,1\}$ uniformly at random and creates the two-qutrit state 
        \begin{equation*}
            \ket{\phi_y} = \frac{1}{\sqrt{2}}\ket{yy} + \frac{1}{\sqrt{2}}\ket{22} \in \A \otimes \B.
        \end{equation*}
        \item Alice sends the qutrit $\mathcal{B}$ to Bob.}\par
        \underline{Stage-II}
        \item Bob chooses $c \in \{0,1\}$ uniformly at random and sends it to Alice.
        \item If $c = 0$, Alice and Bob perform bit commitment as per \cref{protocol:bitCommitment}, otherwise, they perform weak coin flipping as per \cref{protocol:coinFlip:EPR}. 
    \end{itemize}
\end{protocol}

\subsubsection{Cheating Alice}
Dishonest Alice in \cref{protocol:switch:bitCommitment:coinFlip:EPR} would like to maximize her average chances of successfully revealing $\hat{y}$ when Bob decides to perform bit commitment or forcing the outcome $1$ when Bob instead decides to perform weak coin flipping.

The constraints formulated in the SDPs for these two tasks in \ref{sdp:bitCommitment:Alice} and \ref{sdp:coinFlip:EPR:Alice} should be jointly satisfied as Bob can randomly select one of the two tasks. As earlier, we introduce an additional constraint in the form of $\sigma^{BC} = \sigma^{WCF}$ to depict the fact that the message sent by Alice in the beginning of the protocol remains independent of Bob's protocol selection. We finally maximize the average success of Alice for the two tasks under the above constraints.

\textbf{SDP for cheating Alice.} The optimal cheating probability for Alice is given by the optimal objective function value of the following SDP:
\begin{flalign*}
    \text{maximize: } \frac{1}{2}\bigg(\frac{1}{2}\inner{\sigma_0^{BC}}{\kb{\phi_0}} + \frac{1}{2}\inner{\sigma_1^{BC}}{\kb{\phi_1}}\bigg) + \frac{1}{2}\bigg(\frac{1}{2}\inner{\sigma_0^{WCF}}{P_0} + \frac{1}{2}\inner{\sigma_1^{WCF}}{P_1}\bigg) &&
\end{flalign*}
subject to: \\
\begin{minipage}{0.50\linewidth}
    \begin{equation*}
        \begin{aligned}
            & \enspace \Tr_{\A}(\sigma_{\hat{y}}^{BC}) = \sigma^{BC}, \: \forall {\hat{y}} \in \{0,1\} \\
            & \enspace \sigma_{\hat{y}}^{BC} \in \D(\A \otimes \B), \: \forall {\hat{y}} \in \{0,1\} \\
            & \enspace \sigma^{BC} \in \D(\B)       
        \end{aligned}
    \end{equation*}
\end{minipage}%
\begin{minipage}{0.50\linewidth}
\begin{equation*}
    \begin{aligned}
        & \enspace \Tr_{\mathcal{Y} \A_0 \A_1}(\sigma_0^{WCF}) = \Tr_{\mathcal{Y} \A_0 \A_1}(\sigma_1^{WCF}) = \sigma^{WCF}_{B}\\
        & \enspace \Tr_{\B_1}(\sigma^{WCF}_{B}) = \sigma^{WCF} \\
        & \enspace \sigma_0^{WCF}, \sigma_1^{WCF} \in \D(\mathcal{Y} \otimes \A_0 \otimes \B_0 \otimes \A_1 \otimes \B_1) \\
        & \enspace \sigma^{WCF}_{B} \in \D(\B_0 \otimes \B_1) \\
        & \enspace \sigma^{WCF} \in \D(\B_0)
    \end{aligned}
\end{equation*}
\end{minipage}
\begin{equation*}
    \boxed{\sigma^{BC} = \sigma^{WCF}},
\end{equation*}
where $\ket{\phi_0}$ and $\ket{\phi_1}$ are defined in \cref{sdp:bitCommitment:Alice} and $P_0$ and $P_1$ are as defined in \cref{sdp:coinFlip:EPR:Alice}. \\

Numerically solving the above SDP, we find that Alice can cheat with an optimal probability of $P_A = 0.743818$ and this is achieved when $\sigma^{BC} = \sigma^{WCF} = 
\begin{bmatrix}
    0.1281 & 0  &  0 \\
    0 & 0.1281 & 0 \\
    0 & 0 & 0.7438
\end{bmatrix}
$.
Again, note that $P_A < P_A^{BC}$ and $P_A < P_A^{WCF}$ implying an improvement in the security of the switch protocol compared to just \cref{protocol:bitCommitment} or \cref{protocol:coinFlip:EPR}.

\subsubsection{Cheating Bob}

Dishonest Bob would like to maximize the probability of either correctly guessing Alice's commit bit (for bit commitment) or successfully forcing the outcome $0$ (for weak coin flipping). 
In order to systematically analyze the extent of Bob's cheating, we provide a slight modification of~\cref{protocol:switch:bitCommitment:coinFlip:EPR} to simplify Bob's analysis.
Once Alice sends the qutrit $\B_0$, she receives $c$ and the guess $g$ for $y$. She next sends Bob the qutrit $\B_1$ and receives $z$ to obtain protocol outcome here by measuring the qutrit $\A_{\overline{z}}$. 
Note that this modification does not affect the probability of Bob successfully guessing $y$ or forcing the outcome $0$ from the actual protocol and makes the analysis for cheating Bob easier to analyze. This is because Bob sends back his guess $g$ before $\B_1$ is received and furthermore, no additional information is received by Bob in this modified protocol to successfully force the outcome $0$.

Mathematically, consider the state with Alice $\ket{\psi} = \sum \limits_{y} \frac{1}{\sqrt{2}}\ket{y}_{\mathcal{Y}} \ket{\phi_y}_{\A_0 \otimes \B_0} \ket{\phi_y}_{\A_1 \otimes \B_1}$ at the beginning of the protocol. After sending the qutrit $\B_0$, Alice receives $c$ and $g$ from Bob and now holds the state $\tau_0 \in \D(\mathcal{C} \otimes \mathcal{Y} \otimes \A_0 \otimes \A_1 \otimes \B_1 \otimes \mathcal{G})$. She subsequently sends Bob the qutrit $\B_1$ and receives $z$. 
Alice measures her state $\tau \in \D(\mathcal{C} \otimes \mathcal{Y} \otimes \A_0 \otimes \A_1 \otimes \mathcal{Z} \otimes \mathcal{G})$ at the end of the protocol where she accepts, if Bob successfully guesses $y$, or outputs $0$, if Bob successfully forces $0$.
The joint probability of Bob selecting $c=0$ and successfully guessing $y$ is given by $\inner{\tau}{Q_0}$ where
\begin{equation*}
    Q_0 = \kb{0}_{\mathcal{C}} \otimes \Big( \sum_{(y,g) \in \{0,1\}^2} \delta_{y,g} \otimes \kb{y}_{\mathcal{Y}} \otimes \mathbbm{1}_{\A_0} \otimes \mathbbm{1}_{\A_1} \otimes \mathbbm{1}_{\mathcal{Z}} \otimes \kb{g}_{\mathcal{G}} \Big).
\end{equation*}
Similarly, the joint probability of Bob selecting $c=1$ and successfully forcing the outcome $0$ is given by $\inner{\tau}{Q_1}$
where
\begin{equation*}
    Q_1 = \kb{1}_{\mathcal{C}} \otimes \Big( \sum_{z \in \{0,1\}} \mathbbm{1}_{\mathcal{Y}} \otimes (\kb{0} + \kb{1})_{\A_{\overline{z}}} \otimes \mathbbm{1}_{\A_{z}} \otimes \kb{z}_{\mathcal{Z}} \otimes \mathbbm{1}_{\mathcal{G}} \Big).
\end{equation*}
We impose the constraint $ \Tr_{\mathcal{Z}}(\tau) = \Tr_{\B_1}(\tau_0)$ which ensures that the guess state $\mathcal{G}$ in $\tau$ remains the same at the end of the protocol and also the constraint $\Tr_{\mathcal{C} \mathcal{G}}(\tau_0) = \Tr_{\B_0}(\kb{\psi})$. The previous two constraints together ensure that the state on $\A_0,\A_1$ and $\mathcal{Y}$ remains unchanged until the end of the protocol. We finally maximize $\inner{\tau}{Q_0 + Q_1}$ to evaluate the maximum cheating probability of Bob in \cref{protocol:switch:bitCommitment:coinFlip:EPR}. 

\textbf{SDP for cheating Bob.} The optimal cheating probability for Bob is given by the optimal objective function value of the following SDP:
\begin{equation}\label{sdp:switch:bitCommitment:coinFlip:Bob}
    \begin{aligned}
        \text{maximize:} & \enspace \inner{\tau}{Q}  \\
        \text{subject to:} & \enspace \Tr_{\mathcal{C} \mathcal{G}}(\tau_0) = \Tr_{\B_0}(\kb{\psi}) \\
        & \enspace \Tr_{\mathcal{Z}}(\tau) = \Tr_{\B_1}(\tau_0) \\
        & \enspace \tau_0 \in \D(\mathcal{C} \otimes \mathcal{Y} \otimes \A_0 \otimes \A_1 \otimes \B_1 \otimes \mathcal{G}) \\
        & \enspace \tau \in \D(\mathcal{C} \otimes \mathcal{Y} \otimes \A_0 \otimes \A_1 \otimes \mathcal{Z} \otimes \mathcal{G})
    \end{aligned}
\end{equation}
where, 
\begin{equation*}
    \begin{aligned}
        \ket{\psi} &= \sum_{y \in \{0,1\}} \frac{1}{\sqrt{2}}\ket{y}_{\mathcal{Y}} \ket{\phi_y}_{\A_0 \otimes \B_0} \ket{\phi_y}_{\A_1 \otimes \B_1}\\
        Q &= \kb{0}_{\mathcal{C}} \otimes \Big( \sum_{(y,g) \in \{0,1\}^2} \delta_{y,g} \otimes \kb{y}_{\mathcal{Y}} \otimes \mathbbm{1}_{\A_0} \otimes \mathbbm{1}_{\A_1} \otimes \mathbbm{1}_{\mathcal{Z}} \otimes \kb{g}_{\mathcal{G}} \Big)\\
        & \quad + \kb{1}_{\mathcal{C}} \otimes \Big( \sum_{z \in \{0,1\}} \mathbbm{1}_{\mathcal{Y}} \otimes (\kb{0} + \kb{1})_{\A_{\overline{z}}} \otimes \mathbbm{1}_{\A_{z}} \otimes \kb{z}_{\mathcal{Z}} \otimes \mathbbm{1}_{\mathcal{G}} \Big).
    \end{aligned}
\end{equation*}

Numerically solving the above SDP, we find that Bob can cheat with an optimal probability of $P_B = 0.75$.

\subsection{Stochastic switch between oblivious transfer and weak coin flipping}

We present the switch protocol below.
\begin{protocol}{Quantum OT-WCF stochastic switch protocol.}\label{protocol:switch:obliviousTransfer:1outOf2:coinFlip:EPR}
    \begin{itemize}
        \underline{Stage-I}
        \item \textbf{Alice chooses a bit $y \in \{0,1\}$ uniformly at random and creates the two-qutrit state 
        \begin{equation*}
            \ket{\phi_y} = \frac{1}{\sqrt{2}}\ket{yy} + \frac{1}{\sqrt{2}}\ket{22} \in \A \otimes \B.
        \end{equation*}
        \item Alice sends the qutrit $\mathcal{B}$ to Bob.}\par \newpage
        \underline{Stage-II}
        \item Bob chooses $c \in \{0,1\}$ uniformly at random and sends it to Alice.
        \item If $c = 0$, Alice and Bob perform $1$-out-of-$2$ oblivious transfer as per \cref{protocol:obliviousTransfer:1outOf2} , otherwise, they perform weak coin flipping as per \cref{protocol:coinFlip:EPR}.
    \end{itemize}
\end{protocol}

\subsubsection{Cheating Alice}
A dishonest Alice in \cref{protocol:switch:obliviousTransfer:1outOf2:coinFlip:EPR} would like to maximize her average chances of correctly guessing $(x_0,x_1)$  or forcing the outcome $1$ when Bob decides to perform oblivious transfer or weak coin flipping (with equal probability) respectively.

The constraints formulated in the SDP for these two tasks in \ref{sdp:obliviousTransfer:1outOf2:Alice} and \ref{sdp:coinFlip:EPR:Alice} should be jointly satisfied as Bob can randomly select one of the two tasks. As for the SDP of the previous two switch protocols, we introduce an additional constraint in the form of $\sigma_{B}^{OT} = \sigma^{WCF}$ to depict the fact that the message sent by Alice in the beginning of the protocol remains independent of Bob's protocol selection based on the bit $c$. We finally maximize the average success of Alice for the two tasks under the above constraints.

\textbf{SDP for cheating Alice.} The optimal cheating probability for dishonest Alice is given by the optimal objective function value of the following SDP:
\begin{flalign*}
    \text{maximize: } \frac{1}{2}\inner{\sigma^{OT}}{P_{accept}} + \frac{1}{2}\bigg(\frac{1}{2}\inner{\sigma_0^{WCF}}{P_0} + \frac{1}{2}\inner{\sigma_1^{WCF}}{P_1}\bigg) &&
\end{flalign*}
subject to: \\
\begin{minipage}{0.50\linewidth}
    \begin{equation*}
        \begin{aligned}
            & \Tr_{\mathcal{G}_0\mathcal{G}_1}(\sigma^{OT}) = \Tr_{\B}\Big(U_{2}(\kb{\psi} \otimes \sigma_{B}^{OT})U_{2}^{\dagger}\Big) \\
            & \sigma^{OT} \in \D(\mathcal{X}_0 \otimes \mathcal{X}_1 \otimes \mathcal{G}_0 \otimes \mathcal{G}_1) \\
            & \sigma_{B}^{OT} \in \D(\B) \\
        \end{aligned}
    \end{equation*}
\end{minipage}%
\begin{minipage}{0.50\linewidth}
\begin{equation*}
    \begin{aligned}
        & \enspace \Tr_{\mathcal{Y} \A_0 \A_1}(\sigma_0^{WCF}) = \Tr_{\mathcal{Y} \A_0 \A_1}(\sigma_1^{WCF}) = \sigma^{WCF}_{B}\\
        & \enspace \Tr_{\B_1}(\sigma^{WCF}_{B}) = \sigma^{WCF} \\
        & \enspace \sigma_0^{WCF}, \sigma_1^{WCF} \in \D(\mathcal{Y} \otimes \A_0 \otimes \B_0 \otimes \A_1 \otimes \B_1) \\
        & \enspace \sigma^{WCF}_{B} \in \D(\B_0 \otimes \B_1) \\
        & \enspace \sigma^{WCF} \in \D(\B_0)
    \end{aligned}
\end{equation*}
\end{minipage}
\begin{equation*}
    \boxed{\sigma_{B}^{OT} = \sigma^{WCF}},
\end{equation*}
where $P_{accept}$, $U_2$ and $\ket{\psi}$ are defined in \cref{sdp:obliviousTransfer:1outOf2:Alice} and $P_0$ and $P_1$ are as defined in \cref{sdp:coinFlip:EPR:Alice}. \\

Numerically solving the above SDP, we find that Alice can cheat with an optimal probability of $P_A = 0.704407$ and this is achieved when $\sigma_{B}^{OT} = \sigma^{WCF} =
\begin{bmatrix}
    0.22  & 0 &  0 \\
    0  &  0.22  &  0 \\
    0  &  0  &  0.56
\end{bmatrix}$. Again, Alice cheating in the switch protocol gets reduced when compared to its constituent protocols.

\subsubsection{Cheating Bob}

As the objective of dishonest Bob is identical to the formulation of cheating Bob in \cref{protocol:switch:bitCommitment:coinFlip:EPR}, the optimal cheating probability is $P_B^* = 3/4$.

\subsection{Stochastic switch between bit commitment, weak coin flipping, and oblivious transfer}

We present the switch protocol below.
\begin{protocol}{Quantum BC-WCF-OT stochastic switch protocol.}\label{protocol:switch:bitCommitment:obliviousTransfer:coinFlip}
    \begin{itemize}
        \underline{Stage-I}
        \item \textbf{Alice chooses a bit $y \in \{0,1\}$ uniformly at random and creates the two-qutrit state 
        \begin{equation*}
            \ket{\phi_y} = \frac{1}{\sqrt{2}}\ket{yy} + \frac{1}{\sqrt{2}}\ket{22} \in \A \otimes \B.
        \end{equation*}
        \item Alice sends the qutrit $\mathcal{B}$ to Bob.}\par
        \underline{Stage-II}
        \item Bob chooses $c \in \{0,1,2\}$ uniformly at random and sends it to Alice.
        \item If $c = 0$, Alice and Bob perform bit commitment as per \cref{protocol:bitCommitment}, if $c = 1$, they perform weak coin flipping as per \cref{protocol:coinFlip:EPR}, otherwise they perform $1$-out-of-$2$ oblivious transfer as per \cref{protocol:obliviousTransfer:1outOf2}.    
    \end{itemize}
\end{protocol}

\subsubsection{Cheating Alice}

As in the previous switch protocols, a dishonest Alice wishes to maximize her average chances of either successfully revealing $\hat{y}$ (bit commitment), forcing the outcome $1$ (weak coin flipping), or correctly guessing $(x_0,x_1)$ (oblivious transfer) when Bob decides one of these three tasks (with equal probability).

The constraints formulated in the SDPs for the three tasks in \ref{sdp:bitCommitment:Alice}, \ref{sdp:coinFlip:EPR:Alice} and \ref{sdp:obliviousTransfer:1outOf2:Alice} are to be satisfied simultaneously as Bob can randomly select one of the three tasks. We introduce an additional connecting constraint $\sigma^{BC} = \sigma^{WCF} = \sigma_{B}^{OT}$ due to the fact that the qutrit $\B$ is sent by Alice before Bob makes a protocol selection based on his choice $c$. We finally maximize the average success of Alice for the three tasks under the above constraints. 

\textbf{SDP for cheating Alice.} The optimal cheating probability for Alice is given by the optimal objective function value of the following SDP:
\newpage
\begin{flalign*}
    \text{maximize: } \frac{1}{3}\bigg(\sum_{y \in \{0,1\}}\frac{1}{2}\inner{\sigma^{BC}_y}{\kb{\phi_y}}\bigg) + \frac{1}{3}\bigg(\frac{1}{2}\inner{\sigma_0^{WCF}}{P_0} + \frac{1}{2}\inner{\sigma_1^{WCF}}{P_1}\bigg) + \frac{1}{3}\inner{\sigma^{OT}}{P_{accept}} &&
\end{flalign*}
subject to: \\*
\begin{minipage}{0.34\linewidth}
    \begin{equation*}
    \hspace{-2cm}
        \begin{aligned}
            & \enspace \Tr_{\A}(\sigma_{0}^{BC}) = \sigma^{BC} \\
            & \enspace \Tr_{\A}(\sigma_{1}^{BC}) = \sigma^{BC} \\
            & \enspace \sigma_{0}^{BC} \in \D(\A \B), \\
            & \enspace \sigma_{1}^{BC} \in \D(\A \B), \\
            & \enspace \sigma^{BC} \in \D(\B) \\     
        \end{aligned}
    \end{equation*}
\end{minipage}%
\begin{minipage}{0.34\linewidth}
\begin{equation*}
    \hspace{-2cm}
    \begin{aligned}
        & \enspace \Tr_{\mathcal{Y} \A_0 \A_1}(\sigma_0^{WCF}) = \Tr_{\mathcal{Y} \A_0 \A_1}(\sigma_1^{WCF}) \\
        & \enspace \Tr_{\mathcal{Y} \A_0 \A_1}(\sigma_1^{WCF}) = \sigma^{WCF}_{B} \\
        & \enspace \Tr_{\B_1}(\sigma^{WCF}_{B}) = \sigma^{WCF} \\
        & \enspace \sigma_0^{WCF} \in \D(\mathcal{Y} \A_0 \B_0 \A_1  \B_1) \\
        & \enspace \sigma_1^{WCF} \in \D(\mathcal{Y} \A_0 \B_0 \A_1  \B_1) \\
        & \enspace \sigma^{WCF}_{B} \in \D(\B_0 \B_1) \\
        & \enspace \sigma^{WCF} \in \D(\B_0)
    \end{aligned}
\end{equation*}
\end{minipage}
\begin{minipage}{0.33\linewidth}
    \begin{equation*}
        \hspace{-1cm}
        \begin{aligned}
            & \Tr_{\mathcal{G}_0\mathcal{G}_1}(\sigma^{OT}) = \Tr_{\B}\Big(U_{2}(\kb{\psi} \otimes \sigma_{B}^{OT})U_{2}^{\dagger}\Big) \\
            & \sigma^{OT} \in \D(\mathcal{X}_0 \mathcal{X}_1 \mathcal{G}_0 \mathcal{G}_1) \\
            & \sigma_{B}^{OT} \in \D(\B) \\
        \end{aligned}
    \end{equation*}
\end{minipage}%
\begin{equation*}
    \hspace{-3.5cm}
    \boxed{\sigma^{BC} = \sigma^{WCF} = \sigma_{B}^{OT}},
\end{equation*}
where $\ket{\phi_0}$ and $\ket{\phi_1}$ are defined in \cref{sdp:bitCommitment:Alice}, $P_0$ and $P_1$ are defined in \cref{sdp:coinFlip:EPR:Alice}, and $P_{accept}$, $U_2$ and $\ket{\psi}$ are as defined in \cref{sdp:obliviousTransfer:1outOf2:Alice}. \\

Numerically solving the above SDP, we find that Alice can cheat with an optimal probability of $P_A = 0.717779$ and this is achieved when $\sigma^{BC} = \sigma^{WCF} = \sigma^{OT}_{B} = 
\begin{bmatrix}
    0.1987 & 0 & 0 \\
    0 & 0.1987 & 0 \\
    0 & 0 & 0.6026
\end{bmatrix}
$. Note that $P_A$ for the switch between protocols for bit commitment, weak coin flipping, and oblivious transfer is smaller that the cheating probabilities of its constituent protocols. 

\subsubsection{Cheating Bob}
Dishonest Bob would like to maximize the combined probability of correctly guessing $y$ (for bit commitment), successfully forcing the outcome $0$ (for weak coin flipping) or again guessing $y$ correctly (for oblivious transfer). Note that in this switch, $\ket{0}_{\mathcal{C}}$ and $\ket{2}_{\mathcal{C}}$ correspond to Bob selecting bit commitment and oblivious transfer respectively where he tries to correctly guess $y$, while $\ket{1}_{\mathcal{C}}$ corresponds to Bob selecting weak coin flipping where he wishes to force the outcome $0$. The SDP for this three-task switch protocol can thus be formulated along the lines of~\ref{sdp:switch:bitCommitment:coinFlip:Bob} and is described next.

\textbf{SDP for cheating Bob.} The optimal cheating probability for Bob is given by the optimal objective function value of the following SDP:
\begin{equation}\label{sdp:switch:bitCommitment:obliviousTransfer:coinFlip:Bob}
    \begin{aligned}
        \text{maximize:} & \enspace \inner{\tau}{Q}  \\
        \text{subject to:} & \enspace \Tr_{\mathcal{C} \mathcal{G}}(\tau_0) = \Tr_{\B_0}(\kb{\psi}) \\
        & \enspace \Tr_{\mathcal{Z}}(\tau) = \Tr_{\B_1}(\tau_0) \\
        & \enspace \tau_0 \in \D(\mathcal{C} \otimes \mathcal{Y} \otimes \A_0 \otimes \A_1 \otimes \B_1 \otimes \mathcal{G}) \\
        & \enspace \tau \in \D(\mathcal{C} \otimes \mathcal{Y} \otimes \A_0 \otimes \A_1 \otimes \mathcal{Z} \otimes \mathcal{G})
    \end{aligned}
\end{equation}
where, 
\begin{equation*}
    \begin{aligned}
        \ket{\psi} &= \sum_{y \in \{0,1\}} \frac{1}{\sqrt{2}}\ket{y}_{\mathcal{Y}} \ket{\phi_y}_{\A_0 \otimes \B_0} \ket{\phi_y}_{\A_1 \otimes \B_1}\\
        Q &= (\kb{0} + \kb{2})_{\mathcal{C}} \otimes \Big( \sum_{(y,g) \in \{0,1\}^2} \delta_{y,g} \otimes \kb{y}_{\mathcal{Y}} \otimes \mathbbm{1}_{\A_0} \otimes \mathbbm{1}_{\A_1} \otimes \mathbbm{1}_{\mathcal{Z}} \otimes \kb{g}_{\mathcal{G}} \Big)\\
        & \quad + \kb{1}_{\mathcal{C}} \otimes \Big( \sum_{z \in \{0,1\}} \mathbbm{1}_{\mathcal{Y}} \otimes (\kb{0} + \kb{1})_{\A_{\overline{z}}} \otimes \mathbbm{1}_{\A_{z}} \otimes \kb{z}_{\mathcal{Z}} \otimes \mathbbm{1}_{\mathcal{G}} \Big).
    \end{aligned}
\end{equation*}
Numerically solving the above SDP, we find that Bob can cheat with an optimal probability of $P_B = 3/4$.


\section{A quantum protocol for Rabin oblivious transfer} \label{section:depth} 

In the previous section, we noticed that the random subtask selection by Bob in a switch protocol could potentially limit the chances of successful cheating by Alice when compared to her chances in the constituent protocols. 
Therefore, it seems natural to examine the idea of stochastic switching between different protocols \emph{for the same task}. 
In this section, we devise protocols for two variants of Rabin oblivious transfer and in each, we switch between a pair of unsecure protocols to find one with an improved security. 

\subsection{Rabin oblivious transfer}

Rabin oblivious transfer (ROT) is the cryptographic task between two parties, Alice and Bob, where Alice sends a bit $y \in \{0,1\}$ to Bob which he receives with probability $1/2$ and with probability $1/2$ he receives $\perp$ indicating the bit was lost. 
  
We formally define the following notions of security for a given ROT protocol. 
\begin{itemize}
    \item \emph{Completeness:} If both Alice and Bob are honest, then neither party aborts, Bob receives $y$ with probability $1/2$ and $\perp$ with probability $1/2$.
    \item \emph{Cheating Alice} (variant 1):  If Bob is honest, then dishonest Alice's cheating probability is defined as
    \begin{equation*}
        P_A^{ROT1} = \max \Pr[\text{Alice correctly guesses whether Bob received $y$ or learnt $\perp$}]
    \end{equation*}
    where the maximum is taken over all cheating strategies of Alice. Note that $P_A^{ROT1} \geq 1/2$ as Alice can simply choose to follow the protocol honestly and make a uniformly random guess to whether Bob received $y$ or $\perp$ and succeed with probability $1/2$.
    \item \emph{Cheating Alice} (variant 2): If Bob is honest, then dishonest Alice's cheating probability is defined as
    \begin{equation*}
        P_A^{ROT2} = \max \Pr[\text{Alice successfully forces Bob to receive} \perp]
    \end{equation*}
    where the maximum is taken over all cheating strategies of Alice. We have $P_A^{ROT2} \geq 1/2$ as Alice can simply choose to follow the protocol honestly and force Bob to observe $y$ or $\perp$ with probability $1/2$ each.
    \item \emph{Cheating Bob:} If Alice is honest, then dishonest Bob's cheating probability is defined as
    \begin{equation*}
        P_B^{ROT} = \max \Pr[\text{Bob successfully learns $y$}]
    \end{equation*}
    where the maximum is taken over all cheating strategies of Bob. 
Note that $P_B^{ROT} \geq 3/4$ as Bob can simply choose to follow the protocol to learn $y$ with probability $1/2$ and randomly guess $y$ in the event of getting $\perp$.
\end{itemize}

To date, there is little known about the limits of Rabin oblivious transfer and some of the difficulty of defining the security of ROT protocols is in part due to vague cheating desire of Alice. Moreover, Shor's factoring algorithm~\cite{shor1999polynomial} breaks the security of a classical ROT protocol~\cite{fischer1996secure} that relied on the hardness of factoring hinting at the need for developing secure quantum ROT protocols. As far as we are aware, there are no quantum protocols known for ROT in the literature. In this section, we exhibit protocols for the two variants of ROT (defined previously) with poor security which we stochastically combine to construct quantum ROT protocols with improved security.

From here on, we use the notation $\ket{\perp}$ to mean the computational basis state $\ket{2}$ for a qutrit system. 

\subsection{Stochastic Rabin OT switch (variant 1).}

We now present a couple of insecure ROT protocols based on the first variant where Alice wishes to successfully guess whether Bob received the data or $\perp$ and construct another ROT protocol that stochastically switches between these insecure protocols. We further analyze the security of our switch protocol using semidefinite programming to depict a substantial improvement over the security of the constituent protocols.

\begin{protocol}{A simple ROT protocol (variant 1).}\label{protocol:ROT:measure}
    \begin{itemize}
        \underline{Stage-I}
        \item \textbf{Alice chooses a bit $y \in \{0,1\}$ uniformly at random and creates the single-qutrit state 
        \begin{equation*}
            \ket{\phi_y} = \frac{1}{\sqrt{2}}\ket{y} + \frac{1}{\sqrt{2}}\ket{\perp} \in \B.
        \end{equation*}
        \item Alice sends the qutrit $\mathcal{B}$ to Bob.}\par
        \underline{Stage-II}
        \item Bob measures the qutrit $\B$ in computational basis to output $y$ or $\perp$.
    \end{itemize}
\end{protocol}

The above protocol is trivially \emph{complete} as Bob outputs $y$ or $\perp$ each with probability $1/2$ whenever both are honest. However, dishonest Alice can cheat perfectly by sending the qutrit $\kb{\perp}$ to Bob where he outputs $\perp$ with probability $1$. Alternatively, dishonest Bob can perform state discrimination on the two possible states to learn $y$ with a maximum probability $0.9330$.   

\begin{protocol}{A simple ROT protocol with verification (variant 1).}\label{protocol:ROT:verifyAndMeasure}
    \begin{itemize}
        \underline{Stage-I}
        \item \textbf{Alice chooses a bit $y \in \{0,1\}$ uniformly at random and creates the single-qutrit state 
        \begin{equation*}
            \ket{\phi_{y}} = \frac{1}{\sqrt{2}}\ket{y} + \frac{1}{\sqrt{2}}\ket{\perp} \in \B.
        \end{equation*}
        \item Alice sends the qutrit $\mathcal{B}$ to Bob.}\par \newpage
        \underline{Stage-II}
        \item Alice reveals $y$ to Bob where he measures the qutrit state in $\B$ to accept or reject with the POVM:
        \begin{equation*}
            \{\Pi_{accept} \coloneqq \kb{\phi_y}_{\B}, \Pi_{reject} \coloneqq \mathbbm{1}_{\B} - \Pi_{accept}\}.
        \end{equation*}
        If Bob accepts, Alice and Bob restart with \cref{protocol:ROT:measure} (noting the restarted protocol will have a new choice of $y$ for Alice).
    \end{itemize}
\end{protocol}

Dishonest Alice in \cref{protocol:ROT:verifyAndMeasure} could cheat perfectly by not deviating from the protocol until Bob \emph{accepts} and subsequently sends $\kb{\perp}$ once they restart \cref{protocol:ROT:measure}. 
As the first choice of $y$ is completely inconsequential and does not affect the outcome, dishonest Bob would always \emph{accept} and perform state discrimination to successfully learn the (newly chosen) bit $y$ with maximum probability $0.9330$. 

\begin{protocol}{A quantum protocol for ROT based on stochastic switch (variant 1).}\label{protocol:switchROT:verifyBob}
    \begin{itemize}
        \underline{Stage-I}
        \item \textbf{Alice chooses a bit $y_0 \in \{0,1\}$ uniformly at random and creates the single-qutrit state 
        \begin{equation*}
            \ket{\phi_{y_0}} = \frac{1}{\sqrt{2}}\ket{y_0} + \frac{1}{\sqrt{2}}\ket{\perp} \in \B_0.
        \end{equation*}
        \item Alice sends the qutrit $\mathcal{B}_0$ to Bob.}\par
        \underline{Stage-II}
        \item Bob chooses $c \in \{0,1\}$ uniformly at random and sends it to Alice.
        \begin{itemize}
            \item If $c = 0$, Bob measures the qutrit $\B_0$ in computational basis to output $y_0$ or $\perp$.
            \item If $c = 1$, Alice reveals $y_0$ to Bob where he measures $\B_0$ to accept or reject according to the POVM:
            \begin{equation*}
                \{\Pi_{accept} \coloneqq \kb{\phi_{y_0}}_{\B_0}, \Pi_{reject} \coloneqq \mathbbm{1}_{\B_0} - \Pi_{accept}\}.
            \end{equation*}
            If Bob accepts,
                \begin{itemize}
                    \item Alice chooses another bit $y_1 \in \{0,1\}$ uniformly at random and creates the state
                    \begin{equation*}
                        \ket{\phi_{y_1}} = \frac{1}{\sqrt{2}}\ket{y_1} + \frac{1}{\sqrt{2}}\ket{\perp} \in \B_1.
                    \end{equation*}
                    \item Alice sends the qutrit $\B_1$ to Bob.
                    \item Bob measures the qutrit $\B_1$ in the computational basis to output $y_1$ or $\perp$.
                \end{itemize}              
        \end{itemize}
    \end{itemize}
\end{protocol}

\begin{figure}[hbt!]
\centering
\begin{tikzpicture}[x=0.75pt,y=0.75pt,yscale=-1,xscale=1]
\draw   (290,70) -- (410,70) -- (410,100) -- (290,100) -- cycle ;
\draw    (380,100) -- (409.48,148.3) ;
\draw [shift={(410.52,150)}, rotate = 238.6] [color={rgb, 255:red, 0; green, 0; blue, 0 }  ][line width=0.75]    (10.93,-4.9) .. controls (6.95,-2.3) and (3.31,-0.67) .. (0,0) .. controls (3.31,0.67) and (6.95,2.3) .. (10.93,4.9)   ;
\draw    (320,100) -- (275.8,188.21) ;
\draw [shift={(274.91,190)}, rotate = 296.61] [color={rgb, 255:red, 0; green, 0; blue, 0 }  ][line width=0.75]    (10.93,-4.9) .. controls (6.95,-2.3) and (3.31,-0.67) .. (0,0) .. controls (3.31,0.67) and (6.95,2.3) .. (10.93,4.9)   ;
\draw   (309.2,10) -- (390.8,10) .. controls (401.4,10) and (410,18.95) .. (410,30) .. controls (410,41.05) and (401.4,50) .. (390.8,50) -- (309.2,50) .. controls (298.6,50) and (290,41.05) .. (290,30) .. controls (290,18.95) and (298.6,10) .. (309.2,10) -- cycle ;
\draw   (410.52,150) -- (444.99,180.61) -- (409.48,210) -- (375.01,179.39) -- cycle ;
\draw   (350.1,265.22) .. controls (350.05,251.41) and (376.87,240.12) .. (410,240) .. controls (443.13,239.88) and (470.03,250.97) .. (470.08,264.78) .. controls (470.13,278.59) and (443.31,289.88) .. (410.18,290) .. controls (377.05,290.12) and (350.15,279.03) .. (350.1,265.22) -- cycle ;
\draw   (215.01,215.22) .. controls (214.96,201.41) and (241.78,190.12) .. (274.91,190) .. controls (308.04,189.88) and (334.94,200.97) .. (334.99,214.78) .. controls (335.04,228.59) and (308.22,239.88) .. (275.09,240) .. controls (241.96,240.12) and (215.06,229.03) .. (215.01,215.22) -- cycle ;
\draw    (409.48,210) -- (409.97,238) ;
\draw [shift={(410,240)}, rotate = 269] [color={rgb, 255:red, 0; green, 0; blue, 0 }  ][line width=0.75]    (10.93,-3.29) .. controls (6.95,-1.4) and (3.31,-0.3) .. (0,0) .. controls (3.31,0.3) and (6.95,1.4) .. (10.93,3.29)   ;
\draw    (350,50) -- (350,68) ;
\draw [shift={(350,70)}, rotate = 270] [color={rgb, 255:red, 0; green, 0; blue, 0 }  ][line width=0.75]    (10.93,-3.29) .. controls (6.95,-1.4) and (3.31,-0.3) .. (0,0) .. controls (3.31,0.3) and (6.95,1.4) .. (10.93,3.29)   ;
\draw    (444.99,180.61) -- (461.66,181.35) ;
\draw [shift={(463.66,181.44)}, rotate = 182.56] [color={rgb, 255:red, 0; green, 0; blue, 0 }  ][line width=0.75]    (10.93,-3.29) .. controls (6.95,-1.4) and (3.31,-0.3) .. (0,0) .. controls (3.31,0.3) and (6.95,1.4) .. (10.93,3.29)   ;

\draw (241,232.4) node [anchor=north west][inner sep=0.75pt]    {$\ \ \ \ \ \ \ \ \ $};
\draw (350,30) node   [align=left] {\begin{minipage}[lt]{73.89pt}\setlength\topsep{0pt}
{\small Start \cref{protocol:ROT:measure}}
\end{minipage}};
\draw (350,85) node   [align=left] {\begin{minipage}[lt]{73.89pt}\setlength\topsep{0pt}
{\small  \hspace{0.5mm} Bob chooses $c$ }
\end{minipage}};
\draw (410,180) node   [align=left] {\begin{minipage}[lt]{23.8pt}\setlength\topsep{0pt}
{\small \hspace{0.5mm} Test}
\end{minipage}};
\draw (371,342.4) node [anchor=north west][inner sep=0.75pt]    {$\ \ \ \ \ \ \ \ \ $};
\draw (414.5,265) node   [align=left] {\begin{minipage}[lt]{73.89pt}\setlength\topsep{0pt}
{\scriptsize  \ Restart \cref{protocol:ROT:measure}}
\end{minipage}};
\draw (275,215) node   [align=left] {\begin{minipage}[lt]{73.89pt}\setlength\topsep{0pt}
{\scriptsize  Continue \cref{protocol:ROT:measure}}
\end{minipage}};
\draw (250,112.4) node [anchor=north west][inner sep=0.75pt]    {$c=0$};
\draw (411,112.4) node [anchor=north west][inner sep=0.75pt]    {$c=1$};
\draw (421,210) node [anchor=north west][inner sep=0.75pt]   [align=left] {{\small accepts}};
\draw (471,172) node [anchor=north west][inner sep=0.75pt]   [align=left] {rejects};
\end{tikzpicture}
\vspace{-1cm}
\caption{A schematic of our Rabin oblivious transfer protocol. Bob can either decide to continue with the now-in-progress  Protocol~\ref{protocol:ROT:measure} or test and restart Protocol~\ref{protocol:ROT:measure} from the beginning.}
\label{fig:switch:ROT}   
\end{figure}
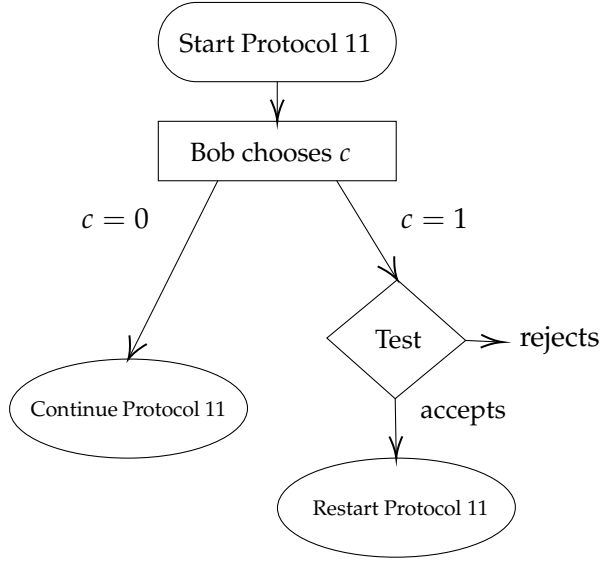

\subsubsection{Cheating Alice (variant 1)}
Dishonest Alice in \cref{protocol:switchROT:verifyBob} would like to maximize her average chances of successfully guessing Bob's outcome ($y$ or $\perp$) when Bob selects $c = 0$ and successfully passing Bob's test when he selects $c = 1$. 
It is worth noting that if Bob accepts, Alice can successfully cheat with probability $1$ in the restarted protocol.

The success probability of Alice can be evaluated by considering the two equally likely scenarios of Bob's selection of $c$. If $\sigma_B$ is the state of the qutrit $\B$ sent by Alice in the first message of the protocol, then the probability with which Alice successfully guesses Bob's outcome when he selects $c = 0$ is given by $\inner{\sigma_0}{P_0}$ where $\sigma_0 \in \D(\mathcal{G} \otimes \B)$ is the state with Bob when Alice sends her guess for the outcome and $P_0$ is given by
\begin{equation*}
    P_0 = \kb{0}_{\mathcal{G}} \otimes (\kb{0} + \kb{1})_{\B} + \kb{1}_{\mathcal{G}} \otimes \kb{\perp}_{\B}
\end{equation*}
is the joint measurement that evaluates whether the guess $\mathcal{G}$ aligns with the measurement outcome. Note that $\ket{0}_{\mathcal{G}}$ means that Alice guesses that Bob has received $y$ and $\ket{1}_{\mathcal{G}}$  means Alice guesses that Bob did not learn anything. 

Alternatively, if $\sigma_1 \in \D(\mathcal{Y} \otimes \B)$ is the state with Bob when he selects $c = 1$, then the joint probability of acceptance by Bob is given by $\inner{\sigma_1}{P_1}$ where,
\begin{equation} 
    P_1 = \sum \limits_{y \in \{0,1\}} \kb{y}_{\mathcal{Y}} \otimes \kb{\phi_y}_{\B}. 
\end{equation} 

\textbf{SDP for cheating Alice.} The optimal cheating probability for Alice is given by the optimal objective function value of the following SDP:
\begin{equation}\label{sdp:ROT:Alice}
    \begin{aligned}
        \text{maximize:} & \enspace \frac{1}{2}\inner{\sigma_0}{P_0} + \frac{1}{2}\inner{\sigma_1}{P_1}  \\
        \text{subject to:} & \enspace \Tr_{\mathcal{G}}(\sigma_0) = \Tr_{\mathcal{Y}}(\sigma_1) = \sigma_B \\
        & \enspace \sigma_0 \in \D(\mathcal{G} \otimes \B) \\
        & \enspace \sigma_1 \in \D(\mathcal{Y} \otimes \B) \\
        & \enspace \sigma_B \in \D(\B)
    \end{aligned}
\end{equation}
where recall,
\begin{equation*}
    \begin{aligned}
        P_0 &= \kb{0}_{\mathcal{G}} \otimes (\kb{0} + \kb{1})_{\B} + \kb{1}_{\mathcal{G}} \otimes \kb{\perp}_{\B} \\
        P_1 &= \sum \limits_{y \in \{0,1\}} \kb{y}_{\mathcal{Y}} \otimes \kb{\phi_y}_{\B}.
    \end{aligned}
\end{equation*}
Numerically solving the above SDP, we find that Alice can cheat with an optimal probability $P_A^{ROT1} = 0.9330$ and this is achieved for $\sigma_{B} = 
\begin{bmatrix}
    0.1890 & -0.1220 & 0.1443 \\
    -0.1220 & 0.1890 & 0.1443 \\
    0.1443 & 0.1443 & 0.6220 
\end{bmatrix}$.

\subsubsection{Cheating Bob}
Dishonest Bob would like to correctly guess $y$, either in the first run of the protocol or the second.
In order to simplify the analysis for cheating Bob, we devise a modified protocol and claim that the extent of Bob's cheating in the modified protocol is same as in the actual protocol. 
For the analysis, consider the state $$\ket{\psi} = \sum \limits_{y_0 \in \{0,1\}} \sum \limits_{y_1 \in \{0,1\}} \frac{1}{2}\ket{y_0}_{\mathcal{Y}_0}\ket{y_0}_{\mathcal{Y}_0'}\ket{y_1}_{\mathcal{Y}_1}\ket{\phi_{y_0}}_{\B_0} \ket{\phi_{y_1}}_{\B_1}$$ with Alice at the beginning of \cref{protocol:switchROT:verifyBob}.
Alice sends Bob the qutrit $\B_0$ while she receives $c$ and a guess $g_0$ for $y_0$. At this point, Alice  holds the state $\tau_0 \in \D(\mathcal{C} \otimes \mathcal{Y}_0 \otimes \mathcal{Y}_0' \otimes \mathcal{Y}_1 \otimes \mathcal{B}_1 \otimes \mathcal{G}_0)$ and sends $y_0$ (in $\mathcal{Y}_0'$) to Bob. Here note that Alice initially held a copy of $y_0$ in $\mathcal{Y}_0'$ which she reveals to Bob and keeps the other to check whether $g_0 = y_0$ at the end of the protocol.
In the final set of messages, Alice sends the qutrit $\B_1$ to Bob and receives $g_1$ as the guess for $y_1$. 
At the end of the protocol, Alice holds the state $\tau \in \D(\mathcal{C} \otimes \mathcal{Y}_0 \otimes \mathcal{Y}_1 \otimes \mathcal{G}_0 \otimes \mathcal{G}_1)$ which she measures to accept either $g_0$, if $c=0$ and Bob correctly guesses $y_0$, or accept $g_1$, if $c=1$ and Bob correctly guesses $y_1$.
The joint probability of Bob selecting $c=0$ and correctly guessing $y_0$ is given by $\inner{\tau}{Q_0}$ where,
\begin{equation*}
    Q_0 = \kb{0}_{\mathcal{C}} \otimes \Big(\sum_{(y_0,g_0) \in \{0,1\}^2} \delta_{y_0,g_0} \kb{y_0}_{\mathcal{Y}_0} \otimes \mathbbm{1}_{\mathcal{Y}_1} \otimes \kb{g_0}_{\mathcal{G}_0} \otimes \mathbbm{1}_{\mathcal{G}_1}\Big).
\end{equation*}
Similarly, the joint probability of Bob selecting $c=1$ and successfully guessing $y_1$ is given by $\inner{\tau}{Q_1}$ where,
\begin{equation*}
Q_1 = \kb{1}_{\mathcal{C}} \otimes \Big(\sum_{(y_1,g_1) \in \{0,1\}^2} \delta_{y_1,g_1} \mathbbm{1}_{\mathcal{Y}_0} \otimes \kb{y_1}_{\mathcal{Y}_1} \otimes \mathbbm{1}_{\mathcal{G}_0} \otimes \kb{g_1}_{\mathcal{G}_1}\Big).
\end{equation*}
To ensure that Bob's guess $(g_0,g_1)$ and Alice's input $(y_0, y_1)$ remains unchanged, we introduce the constraints $\Tr_{\mathcal{Y}_0' \B_1}(\tau_0) = \Tr_{\mathcal{G}_1}(\tau)$ and $\Tr_{\mathcal{C}\mathcal{G}_0}(\tau_0) = \Tr_{\B_0}(\kb{\psi})$. We finally maximize $\inner{\tau}{Q_0 + Q_1}$ to obtain the optimal cheating probability for Bob.

\textbf{SDP for cheating Bob.} The optimal cheating probability for Bob is given by the optimal objective function value of the following SDP:

\begin{equation}\label{sdp:ROT:Bob}
    \begin{aligned}
        \text{maximize:} & \enspace \inner{\tau}{Q} \\
        \text{subject to:} & \enspace \Tr_{\mathcal{C}\mathcal{G}_0}(\tau_0) = \Tr_{\B_0}(\kb{\psi}) \\
        & \enspace \Tr_{\mathcal{Y}_0' \B_1}(\tau_0) = \Tr_{\mathcal{G}_1}(\tau) \\
        & \enspace \tau_0 \in \D(\mathcal{C} \otimes \mathcal{Y}_0 \otimes \mathcal{Y}_0' \otimes \mathcal{Y}_1 \otimes \mathcal{B}_1 \otimes \mathcal{G}_0) \\
        & \enspace \tau \in \D(\mathcal{C} \otimes \mathcal{Y}_0 \otimes \mathcal{Y}_1 \otimes \mathcal{G}_0 \otimes \mathcal{G}_1)
    \end{aligned}
\end{equation}
where recall, 
\begin{equation}
    \begin{aligned}
        \ket{\psi} &= \sum_{y_0 \in \{0,1\}} \sum_{y_1 \in \{0,1\}]}\frac{1}{2}\ket{y_0}_{\mathcal{Y}_0}\ket{y_0}_{\mathcal{Y}_0'}\ket{y_1}_{\mathcal{Y}_1}\ket{\phi_{y_0}}_{\B_0}\ket{\phi_{y_1}}_{\B_1}\\
        Q &= \kb{0}_{\mathcal{C}} \otimes \Big(\sum_{(y_0,g_0) \in \{0,1\}^2} \delta_{y_0,g_0} \kb{y_0}_{\mathcal{Y}_0} \otimes \mathbbm{1}_{\mathcal{Y}_1} \otimes \kb{g_0}_{\mathcal{G}_0} \otimes \mathbbm{1}_{\mathcal{G}_1}\Big) \\
        & \quad + \kb{1}_{\mathcal{C}} \otimes \Big(\sum_{(y_1,g_1) \in \{0,1\}^2} \delta_{y_1,g_1} \mathbbm{1}_{\mathcal{Y}_0} \otimes \kb{y_1}_{\mathcal{Y}_1} \otimes \mathbbm{1}_{\mathcal{G}_0} \otimes \kb{g_1}_{\mathcal{G}_1}\Big).
    \end{aligned}
\end{equation}
Numerically solving the above SDP, we find that Bob can cheat with an optimal probability of $P_B^{ROT} = 0.9691$.

\subsection{Stochastic Rabin OT switch (variant 2)}
Next, we develop and analyze another ROT protocol for the second variant where Alice wishes to force $\perp$. The constant lower bounds for such task are discussed in~\cite{osborn2022constant}. The ROT protocol in discussion also relies on the idea of stochastic switching between insecure protocols to develop another protocol with improved security.

\begin{protocol}{An alternative ROT protocol based on stochastic switch.}\label{protocol:switchROT:verifyBob:alternate}
    \begin{itemize}
        \underline{Stage-I}
        \item \textbf{Alice chooses a bit $y_0 \in \{0,1\}$ uniformly at random and creates the two-qutrit state 
        \begin{equation*}
            \ket{\phi_{y_0}} = \frac{1}{\sqrt{2}}\ket{y_0y_0} + \frac{1}{\sqrt{2}}\ket{\perp \perp} \in \A_0 \otimes \B_0.
        \end{equation*}
        \item Alice sends the qutrit $\B_0$ to Bob.}\par
        \underline{Stage-II}
        \item Bob chooses $c \in \{0,1\}$ uniformly at random and sends it to Alice.
        \begin{itemize}
            \item If $c = 0$, Bob measures the qutrit $\B_0$ in computational basis to output $y_0$ or $\perp$.
            \item If $c = 1$, Alice reveals $y_0$ to Bob and sends him the qutrit $\A_0$ where Bob measures the combined state in $\A_0 \otimes \B_0$ to accept or reject according to the POVM:
            \begin{equation*}
                \{\Pi_{accept} \coloneqq \kb{\phi_{y_0}}_{\A_0 \otimes \B_0}, \Pi_{reject} \coloneqq \mathbbm{1}_{\A_0 \otimes \B_0} - \Pi_{accept}\}.
            \end{equation*}
            If Bob accepts, 
            \begin{itemize}
                \item Alice chooses another bit $y_1 \in \{0,1\}$ uniformly at random and creates the two-qutrit state 
                \begin{equation*}
                    \ket{\phi_{y_1}} = \frac{1}{\sqrt{2}}\ket{y_1y_1} + \frac{1}{\sqrt{2}}\ket{\perp \perp} \in \A_1 \otimes \B_1.
                \end{equation*}
                \item Alice sends the qutrit $\B_1$ to Bob.
                \item Bob measures the qutrit $\B_1$ in computational basis to output $y_1$ or $\perp$.
            \end{itemize}
        \end{itemize}
    \end{itemize}
\end{protocol}

\subsubsection{Cheating Alice (variant 2)}
Dishonest Alice in \cref{protocol:switchROT:verifyBob:alternate} would like to maximize her average chances of successfully forcing $\perp$ when Bob selects $c = 0$ and successfully passing Bob's test when he selects $c = 1$ (note that if Bob accepts, Alice can successfully force $\perp$ with probability $1$ in the restarted protocol). Let $\sigma_B$ be the state with Bob after Alice the qutrit $\B$ in her first message of the protocol. Then $\inner{\sigma_B}{\kb{\perp}}$ is the probability by which Bob observes nothing ($\perp$) if he chooses $c = 0$. On the other hand, if Bob chooses $c=1$, then Alice sends the qutrit $\A$ such that Bob accepts the joint state in $\A \otimes \B$ with a high probability given by $\inner{\sigma}{P}$ where
\begin{equation}
    P = \sum \limits_{y \in \{0,1\}} \kb{y}_{\mathcal{Y}} \otimes \kb{\phi_y}_{\A \otimes \B}.
\end{equation}
Note that Alice can successfully force $\perp$ to Bob with probability $1$ if Alice accepts in the latter case.

\textbf{SDP for cheating Alice.} The optimal cheating probability for Alice is given by the optimal objective function value of the following SDP:
\begin{equation}\label{sdp:ROT:Alice:alternate}
    \begin{aligned}
        \text{maximize:} & \enspace \frac{1}{2}\inner{\sigma_B}{\kb{\perp}} + \frac{1}{2}\inner{\sigma}{P}  \\
        \text{subject to:} & \enspace \Tr_{\mathcal{Y}\A}(\sigma) = \sigma_B \\
        & \enspace \sigma \in \D(\mathcal{Y} \otimes \A \otimes \B) \\
        & \enspace \sigma_{B} \in \D(\B)
    \end{aligned}
\end{equation}
where recall,
\begin{equation*}
    \begin{aligned}
        P &= \sum \limits_{y \in \{0,1\}} \kb{y}_{\mathcal{Y}} \otimes \kb{\phi_y}_{\A \otimes \B}.
    \end{aligned}
\end{equation*}
Numerically solving the above SDP (mapping $\ket{\perp}$ to the computational basis state $\ket{2}$), we find that Alice can cheat with an optimal probability of $P_A^{ROT2} = \cos^2(\pi/8) \approx 0.8535$ and this is achieved for $\sigma_{B} = 
\begin{bmatrix}
    0.4268 & 0 & 0 \\
    0 & 0.4268 & 0 \\
    0 & 0 & 0.1464 
\end{bmatrix}$.
 
\subsubsection{Cheating Bob}

As before, dishonest Bob would like to correctly guess $y$, either in the first run of the protocol or the second.

Consider the state $$\ket{\psi} = \sum_{y_0 \in \{0,1\}} \sum_{y_1 \in \{0,1\}]}\frac{1}{2}\ket{y_0}_{\mathcal{Y}_0}\ket{y_0}_{\mathcal{Y}_0'}\ket{y_1}_{\mathcal{Y}_1}\ket{\phi_{y_0}}_{\A_0 \otimes \B_0}\ket{\phi_{y_1}}_{\A_1 \otimes \B_1}$$ with Alice at the beginning of \cref{protocol:switchROT:verifyBob:alternate}.
Alice sends Bob the qutrit $\B_0$ while she receives $c$ and a guess $g_0$ for $y_0$. At this point, Alice holds the state $\tau_0 \in \D(\mathcal{C} \otimes \mathcal{Y}_0 \otimes \mathcal{Y}_0' \otimes \mathcal{Y}_1 \otimes \mathcal{A}_0 \otimes \mathcal{A}_1 \otimes \mathcal{B}_1 \otimes \mathcal{G}_0)$ and sends a copy of $y_0$ (in $\mathcal{Y}_0'$) and $\A_0$ to Bob. 

In the final set of messages, Alice sends the qutrit $\B_1$ to Bob and receives $g_1$ as the guess for $y_1$. 
Alice holds the state $\tau \in \D(\mathcal{C} \otimes \mathcal{Y}_0 \otimes \mathcal{Y}_1 \otimes \mathcal{A}_1 \otimes \mathcal{G}_0 \otimes \mathcal{G}_1)$ at the end of the protocol and measures to accept $g_0$ (using the remaining copy of $y_0$ in $\mathcal{Y}_0$), if Bob selects $c=0$ and correctly guesses $y_0$, or accept $g_1$, if Bob selects $c=1$ and correctly guesses $y_1$.
The joint probability of Bob selecting $c=0$ and successfully guessing $y_0$ is given by $\inner{\tau}{Q_0}$ where
\begin{equation*}
    Q_0 = \kb{0}_{\mathcal{C}} \otimes \Big(\sum_{(y_0,g_0) \in \{0,1\}^2} \delta_{y_0,g_0} \kb{y_0}_{\mathcal{Y}_0} \otimes \mathbbm{1}_{\mathcal{Y}_1} \otimes \mathbbm{1}_{\A_1} \otimes \kb{g_0}_{\mathcal{G}_0} \otimes \mathbbm{1}_{\mathcal{G}_1}\Big).
\end{equation*}
Similarly, the joint probability of Bob selecting $c=1$ and successfully guessing $y_1$ is given by $\inner{\tau}{Q_1}$ where
\begin{equation*}
Q_1 = \kb{1}_{\mathcal{C}} \otimes \Big(\sum_{(y_1,g_1) \in \{0,1\}^2} \delta_{y_1,g_1} \mathbbm{1}_{\mathcal{Y}_0} \otimes \kb{y_1}_{\mathcal{Y}_1} \otimes \mathbbm{1}_{\A_1} \otimes \mathbbm{1}_{\mathcal{G}_0} \otimes \kb{g_1}_{\mathcal{G}_1}\Big).
\end{equation*}
We introduce the constraints $\Tr_{\mathcal{Y}_0' \A_0 \B_1}(\tau_0) = \Tr_{\mathcal{G}_1}(\tau)$ and $\Tr_{\mathcal{C}\mathcal{G}_0}(\tau_0) = \Tr_{\B_0}(\kb{\psi})$ to ensure that $(g_0,g_1)$ and Alice's input $(y_0, y_1)$ remains unchanged. We finally maximize $\inner{\tau}{Q_0 + Q_1}$ over $\tau$ and $\tau_0$ to obtain the optimal cheating probability for Bob.

\textbf{SDP for cheating Bob.} The optimal cheating probability for Bob is given by the optimal objective function value of the following SDP:
\begin{equation}\label{sdp:ROT:Bob:alternate}
    \begin{aligned}
        \text{maximize:} & \enspace \inner{\tau}{Q} \\
        \text{subject to:} & \enspace \Tr_{\mathcal{C}\mathcal{G}_0}(\tau_0) = \Tr_{\B_0}(\kb{\psi}) \\
        & \enspace \Tr_{\mathcal{Y}_0' \A_0 \B_1}(\tau_0) = \Tr_{\mathcal{G}_1}(\tau) \\
        & \enspace \tau_0 \in \D(\mathcal{C} \otimes \mathcal{Y}_0 \otimes \mathcal{Y}_0' \otimes \mathcal{Y}_1 \otimes \mathcal{A}_0 \otimes \mathcal{A}_1 \otimes \mathcal{B}_1 \otimes \mathcal{G}_0) \\
        & \enspace \tau \in \D(\mathcal{C} \otimes \mathcal{Y}_0 \otimes \mathcal{Y}_1 \otimes \mathcal{A}_1 \otimes \mathcal{G}_0 \otimes \mathcal{G}_1)
    \end{aligned}
\end{equation}
where recall, 
\begin{equation}
    \begin{aligned}
        \ket{\psi} &= \sum_{y_0 \in \{0,1\}} \sum_{y_1 \in \{0,1\}]}\frac{1}{2}\ket{y_0}_{\mathcal{Y}_0}\ket{y_0}_{\mathcal{Y}_0'}\ket{y_1}_{\mathcal{Y}_1}\ket{\phi_{y_0}}_{\A_0 \otimes \B_0}\ket{\phi_{y_1}}_{\A_1 \otimes \B_1}\\
        Q &= \kb{0}_{\mathcal{C}} \otimes \Big(\sum_{(y_0,g_0) \in \{0,1\}^2} \delta_{y_0,g_0} \kb{y_0}_{\mathcal{Y}_0} \otimes \mathbbm{1}_{\mathcal{Y}_1} \otimes \mathbbm{1}_{\A_1} \otimes \kb{g_0}_{\mathcal{G}_0} \otimes \mathbbm{1}_{\mathcal{G}_1}\Big) \\
        & \quad + \kb{1}_{\mathcal{C}} \otimes \Big(\sum_{(y_1,g_1) \in \{0,1\}^2} \delta_{y_1,g_1} \mathbbm{1}_{\mathcal{Y}_0} \otimes \kb{y_1}_{\mathcal{Y}_1} \otimes \mathbbm{1}_{\A_1} \otimes \mathbbm{1}_{\mathcal{G}_0} \otimes \kb{g_1}_{\mathcal{G}_1}\Big).
    \end{aligned}
\end{equation}

Numerically solving the above SDP, we find that Bob can cheat with an optimal probability of $P_B^{ROT} = 7/8$.

It is important to note the difference in nature of state preparations by Alice in the switch protocols for the two variants of the Rabin OT task (\cref{protocol:switchROT:verifyBob} and \cref{protocol:switchROT:verifyBob:alternate}). The single qutrit state prepared by Alice in~\cref{protocol:switchROT:verifyBob} for the first variant helps ensure that dishonest Alice is unable to learn Bob's measurement output perfectly which may not be the case when an entangled state is shared between them. Similarly, Alice sending only a part of an entangled state in her initial message to Bob in~\cref{protocol:switchROT:verifyBob:alternate} lowers the overall probability of Bob successfully guessing Alice's bit when compared to~\cref{protocol:switchROT:verifyBob} where she sends the entire single qutrit state in her first message to Bob.

\section*{Computational platform}
The semidefinite programs formulated in this work were solved using CVX, a package for specifying and solving convex programs~\cite{cvx, gb08}, with the help of the solver MOSEK~\cite{mosek} and supported by the methods of QETLAB~\cite{qetlab}, a MATLAB toolbox for exploring quantum entanglement theory. 

The codes developed to solve the semidefinite programs can be found at the following git repository: \url{https://bitbucket.org/akshaybansal14/two-party-cryptography/}.

\section*{Acknowledgements} 
This work is supported in part by Commonwealth Cyber Initiative SWVA grant 453136. We also thank the Computer Science Research Virtual Machine Project (CSRVM) at Virginia Tech for providing the computational resources for our work.


\nocite{*}
\bibliographystyle{quantum}
\bibliography{references} 

\begin{thebibliography}{10}

\bibitem{wiesner1983conjugate}
Stephen Wiesner.
\newblock ``Conjugate coding''.
\newblock \href{https://dx.doi.org/https://doi.org/10.1145/1008908.1008920}{ACM
  Sigact News {\bf 15}, 78--88}~(1983).

\bibitem{rabin2005exchange}
Michael~O. Rabin.
\newblock ``How to exchange secrets with oblivious transfer''.
\newblock Cryptology {ePrint} Archive, Paper 2005/187~(2005).
\newblock \url{https://eprint.iacr.org/2005/187}.

\bibitem{crepeau1994quantum}
Claude Cr{\'e}peau.
\newblock ``Quantum oblivious transfer''.
\newblock
  \href{https://dx.doi.org/https://doi.org/10.1080/09500349414552291}{Journal
  of Modern Optics {\bf 41}, 2445--2454}~(1994).

\bibitem{osborn2022constant}
Sarah~A. Osborn and Jamie Sikora.
\newblock ``{A Constant Lower Bound for Any Quantum Protocol for Secure
  Function Evaluation}''.
\newblock In Fran\c{c}ois Le~Gall and Tomoyuki Morimae, editors, 17th
  Conference on the Theory of Quantum Computation, Communication and
  Cryptography (TQC 2022).
\newblock
  \href{https://dx.doi.org/https://doi.org/10.4230/LIPIcs.TQC.2022.8}{Volume
  232 of Leibniz International Proceedings in Informatics (LIPIcs), pages
  8:1--8:14}.
\newblock Dagstuhl, Germany~(2022). Schloss Dagstuhl -- Leibniz-Zentrum f{\"u}r
  Informatik.

\bibitem{schaffner2007cryptography}
Schaffner Christian.
\newblock ``Cryptography in the bounded-quantum-storage model''~(2007).
\newblock  \href{http://arxiv.org/abs/0709.0289}{arXiv:0709.0289}.

\bibitem{chailloux2013lowerot}
Andr{\'e} Chailloux, Iordanis Kerenidis, and Jamie Sikora.
\newblock ``Lower bounds for quantum oblivious transfer''.
\newblock
  \href{https://dx.doi.org/https://doi.org/10.26421/QIC13.1-2-9}{Quantum
  Information \& Computation {\bf 13}, 158--177}~(2013).

\bibitem{buhrman2012complete}
Harry Buhrman, Matthias Christandl, and Christian Schaffner.
\newblock ``Complete insecurity of quantum protocols for classical two-party
  computation''.
\newblock
  \href{https://dx.doi.org/https://doi.org/10.1103/PhysRevLett.109.160501}{Physical
  Review Letters {\bf 109}, 160501}~(2012).

\bibitem{amiri2021imperfect}
Ryan Amiri, Robert St{\'a}rek, David Reichmuth, Ittoop~V Puthoor, Michal
  Mi{\v{c}}uda, Ladislav Mi{\v{s}}ta~Jr, Miloslav Du{\v{s}}ek, Petros Wallden,
  and Erika Andersson.
\newblock ``Imperfect 1-out-of-2 quantum oblivious transfer: bounds, a
  protocol, and its experimental implementation''.
\newblock
  \href{https://dx.doi.org/https://doi.org/10.1103/PRXQuantum.2.010335}{PRX
  Quantum {\bf 2}, 010335}~(2021).

\bibitem{chailloux2016semihonest}
Gus~Gutoski Andr\'e~Chailloux and Jamie Sikora.
\newblock ``Optimal bounds for semi-honest quantum oblivious transfer''.
\newblock
  \href{https://dx.doi.org/http://dx.doi.org/10.4086/cjtcs.2016.013}{Chicago
  Journal of Theoretical Computer Science {\bf 13}, 1--17}~(2016).

\bibitem{ambainis2001new}
Andris Ambainis.
\newblock ``A new protocol and lower bounds for quantum coin flipping''.
\newblock In Proceedings of the Thirty-Third Annual ACM Symposium on Theory of
  Computing.
\newblock \href{https://dx.doi.org/10.1145/380752.380788}{Page 134–142}.
\newblock STOC '01New York, NY, USA~(2001). Association for Computing
  Machinery.

\bibitem{mochon2007quantum}
Carlos Mochon.
\newblock ``Quantum weak coin flipping with arbitrarily small bias''~(2007).
\newblock  \href{http://arxiv.org/abs/0711.4114}{arXiv:0711.4114}.

\bibitem{aharonov2016simpler}
Dorit Aharonov, Andr{\'e} Chailloux, Maor Ganz, Iordanis Kerenidis, and
  Lo{\"\i}ck Magnin.
\newblock ``A simpler proof of the existence of quantum weak coin flipping with
  arbitrarily small bias''.
\newblock \href{https://dx.doi.org/https://doi.org/10.1137/14096387X}{SIAM
  Journal on Computing {\bf 45}, 633--679}~(2016).

\bibitem{arora2019quantum}
Atul~Singh Arora, J\'{e}r\'{e}mie Roland, and Stephan Weis.
\newblock ``Quantum weak coin flipping''.
\newblock In Proceedings of the 51st Annual ACM SIGACT Symposium on Theory of
  Computing.
\newblock \href{https://dx.doi.org/10.1145/3313276.3316306}{Page 205–216}.
\newblock STOC 2019New York, NY, USA~(2019). Association for Computing
  Machinery.

\bibitem{arora2021analytic}
Atul~Singh Arora, J{\'e}r{\'e}mie Roland, and Chrysoula Vlachou.
\newblock ``Analytic quantum weak coin flipping protocols with arbitrarily
  small bias''.
\newblock In Proceedings of the 2021 ACM-SIAM Symposium on Discrete Algorithms
  (SODA).
\newblock
  \href{https://dx.doi.org/https://doi.org/10.1137/1.9781611976465.58}{Pages
  919--938}.
\newblock ~(2021).

\bibitem{miller2020impossibility}
Carl~A. Miller.
\newblock ``The impossibility of efficient quantum weak coin flipping''.
\newblock In Proceedings of the 52nd Annual ACM SIGACT Symposium on Theory of
  Computing.
\newblock
  \href{https://dx.doi.org/https://doi.org/10.1145/3357713.3384276}{Page
  916–929}.
\newblock STOC 2020New York, NY, USA~(2020). Association for Computing
  Machinery.

\bibitem{mochon2005large}
Carlos Mochon.
\newblock ``Large family of quantum weak coin-flipping protocols''.
\newblock
  \href{https://dx.doi.org/https://doi.org/10.1103/PhysRevA.72.022341}{Physical
  Review A {\bf 72}, 022341}~(2005).

\bibitem{kitaev2002quantum}
Alexei Kitaev.
\newblock ``Quantum coin flipping''.
\newblock In 6th {A}nnual workshop on {Q}uantum {I}nformation {P}rocessing
  ({U}npublished result).
\newblock ~(2002).

\bibitem{aharon2010quantum}
Nati Aharon and Jonathan Silman.
\newblock ``Quantum dice rolling: a multi-outcome generalization of quantum
  coin flipping''.
\newblock
  \href{https://dx.doi.org/https://doi.org/10.1088/1367-2630/12/3/033027}{New
  Journal of Physics {\bf 12}, 033027}~(2010).

\bibitem{barnum2007generalized}
Howard Barnum, Jonathan Barrett, Matthew Leifer, and Alexander Wilce.
\newblock ``Generalized no-broadcasting theorem''.
\newblock
  \href{https://dx.doi.org/https://doi.org/10.1103/PhysRevLett.99.240501}{Physical
  Review Letters {\bf 99}, 240501}~(2007).

\bibitem{wootters2009no}
WK~Wootters and WH~Zurek.
\newblock ``A single quantum cannot be cloned''.
\newblock \href{https://dx.doi.org/https://doi.org/10.1038/299802a0}{Nature
  {\bf 299}, 802--803}~(1982).

\bibitem{lindell2016fast}
Yehuda Lindell.
\newblock ``Fast cut-and-choose-based protocols for malicious and covert
  adversaries''.
\newblock
  \href{https://dx.doi.org/https://doi.org/10.1007/s00145-015-9198-0}{Journal
  of Cryptology {\bf 29}, 456--490}~(2016).

\bibitem{sturm1999using}
Jos~F Sturm.
\newblock ``Using {S}e{D}u{M}i 1.02, a {MATLAB} toolbox for optimization over
  symmetric cones''.
\newblock
  \href{https://dx.doi.org/https://doi.org/10.1080/10556789908805766}{Optimization
  methods and software {\bf 11}, 625--653}~(1999).

\bibitem{toh1999sdpt3}
Kim-Chuan Toh, Michael~J Todd, and Reha~H T{\"u}t{\"u}nc{\"u}.
\newblock ``{SDPT3}—a {MATLAB} software package for semidefinite programming,
  version 1.3''.
\newblock
  \href{https://dx.doi.org/https://doi.org/10.1080/10556789908805762}{Optimization
  Methods and Software {\bf 11}, 545--581}~(1999).

\bibitem{mosek}
MOSEK ApS.
\newblock ``The mosek optimization toolbox for matlab manual. version 9.0.''.
\newblock ~(2019).
\newblock  url:~\url{http://docs.mosek.com/9.0/toolbox/index.html}.

\bibitem{nayak2003bit}
Ashwin Nayak and Peter Shor.
\newblock ``Bit-commitment-based quantum coin flipping''.
\newblock
  \href{https://dx.doi.org/https://doi.org/10.1103/PhysRevA.67.012304}{Physical
  Review A {\bf 67}, 012304}~(2003).

\bibitem{nielsen2010quantum}
Michael~A Nielsen and Isaac~L Chuang.
\newblock ``Quantum {C}omputation and {Q}uantum {I}nformation''.
\newblock
  \href{https://dx.doi.org/https://doi.org/10.1017/CBO9780511976667}{Cambridge
  University Press}. ~(2010).

\bibitem{watrous2018theory}
John Watrous.
\newblock ``The {T}heory of {Q}uantum {I}nformation''.
\newblock
  \href{https://dx.doi.org/https://doi.org/10.1017/9781316848142}{Cambridge
  University Press}. ~(2018).

\bibitem{helmberg1996interior}
Christoph Helmberg, Franz Rendl, Robert~J Vanderbei, and Henry Wolkowicz.
\newblock ``An interior-point method for semidefinite programming''.
\newblock \href{https://dx.doi.org/https://doi.org/10.1137/0806020}{SIAM
  Journal on Optimization {\bf 6}, 342--361}~(1996).

\bibitem{alizadeh1998primal}
Farid Alizadeh, Jean-Pierre~A Haeberly, and Michael~L Overton.
\newblock ``Primal-dual interior-point methods for semidefinite programming:
  convergence rates, stability and numerical results''.
\newblock
  \href{https://dx.doi.org/https://doi.org/10.1137/S1052623496304700}{SIAM
  Journal on Optimization {\bf 8}, 746--768}~(1998).

\bibitem{boyd2004convex}
Stephen Boyd and Lieven Vandenberghe.
\newblock ``Convex {O}ptimization''.
\newblock
  \href{https://dx.doi.org/https://doi.org/10.1017/CBO9780511804441}{Cambridge
  {U}niversity {P}ress}. ~(2004).

\bibitem{mehrotra2007decomposition}
Sanjay Mehrotra and M~G{\"o}khan {\"O}zevin.
\newblock ``Decomposition-based interior point methods for two-stage stochastic
  semidefinite programming''.
\newblock \href{https://dx.doi.org/https://doi.org/10.1137/050622067}{SIAM
  Journal on Optimization {\bf 18}, 206--222}~(2007).

\bibitem{lo1997quantum}
Hoi-Kwong Lo and Hoi~Fung Chau.
\newblock ``Is quantum bit commitment really possible?''.
\newblock
  \href{https://dx.doi.org/https://doi.org/10.1103/PhysRevLett.78.3410}{Physical
  Review Letters {\bf 78}, 3410}~(1997).

\bibitem{mayers1997unconditionally}
Dominic Mayers.
\newblock ``Unconditionally secure quantum bit commitment is impossible''.
\newblock
  \href{https://dx.doi.org/https://doi.org/10.1103/PhysRevLett.78.3414}{Physical
  Review Letters {\bf 78}, 3414}~(1997).

\bibitem{sikora2018bitcommitment}
Jamie Sikora and John~H Selby.
\newblock ``Simple proof of the impossibility of bit commitment in generalized
  probabilistic theories using cone programming''.
\newblock
  \href{https://dx.doi.org/https://doi.org/10.1103/PhysRevA.97.042302}{Physical
  Review A {\bf 97}, 042302}~(2018).

\bibitem{chailloux2011bc}
Andre Chailloux and Iordanis Kerenidis.
\newblock ``Optimal bounds for quantum bit commitment''.
\newblock In 2011 IEEE 52nd Annual Symposium on Foundations of Computer
  Science.
\newblock \href{https://dx.doi.org/https://doi.org/10.1109/FOCS.2011.42}{Pages
  354--362}.
\newblock ~(2011).

\bibitem{chailloux2009scf}
André Chailloux and Iordanis Kerenidis.
\newblock ``Optimal quantum strong coin flipping''.
\newblock In 2009 50th Annual IEEE Symposium on Foundations of Computer
  Science.
\newblock \href{https://dx.doi.org/https://doi.org/10.1109/FOCS.2009.71}{Pages
  527--533}.
\newblock ~(2009).

\bibitem{kerenidis2004weak}
Iordanis Kerenidis and Ashwin Nayak.
\newblock ``Weak coin flipping with small bias''.
\newblock
  \href{https://dx.doi.org/https://doi.org/10.1016/j.ipl.2003.07.007}{Information
  Processing Letters {\bf 89}, 131--135}~(2004).

\bibitem{colbeck2007entanglement}
Roger Colbeck.
\newblock ``An entanglement-based protocol for strong coin tossing with bias
  1/4''.
\newblock
  \href{https://dx.doi.org/https://doi.org/10.1016/j.physleta.2006.10.062}{Physics
  Letters A {\bf 362}, 390--392}~(2007).

\bibitem{shor1999polynomial}
Peter~W Shor.
\newblock ``Polynomial-time algorithms for prime factorization and discrete
  logarithms on a quantum computer''.
\newblock
  \href{https://dx.doi.org/https://doi.org/10.1137/S0036144598347011}{SIAM
  review {\bf 41}, 303--332}~(1999).

\bibitem{fischer1996secure}
Michael~J Fischer, Silvio Micali, and Charles Rackoff.
\newblock ``A secure protocol for the oblivious transfer''.
\newblock \href{https://dx.doi.org/https://doi.org/10.1007/BF00208002}{Journal
  of Cryptology {\bf 9}, 191--196}~(1996).

\bibitem{cvx}
Michael Grant and Stephen Boyd.
\newblock ``{CVX}: Matlab software for disciplined convex programming, version
  2.1''.
\newblock \url{http://cvxr.com/cvx}~(2014).

\bibitem{gb08}
Michael Grant and Stephen Boyd.
\newblock ``Graph implementations for nonsmooth convex programs''.
\newblock In V.~Blondel, S.~Boyd, and H.~Kimura, editors, Recent Advances in
  Learning and Control.
\newblock
  \href{https://dx.doi.org/https://doi.org/10.1007/978-1-84800-155-8_7}{Pages
  95--110}.
\newblock Lecture Notes in Control and Information Sciences. Springer-Verlag
  Limited~(2008).

\bibitem{qetlab}
Nathaniel Johnston.
\newblock ``{QETLAB}: A {MATLAB} toolbox for quantum entanglement, version
  0.9''.
\newblock \url{http://qetlab.com}~(2016).

\bibitem{aharon2011weak}
Nati Aharon, Andr{\'e} Chailloux, Iordanis Kerenidis, Serge Massar, Stefano
  Pironio, and Jonathan Silman.
\newblock ``Weak coin flipping in a device-independent setting''.
\newblock In Theory of Quantum Computation, Communication, and Cryptography:
  Revised Selected Papers.
\newblock
  \href{https://dx.doi.org/https://doi.org/10.1007/978-3-642-54429-3_1}{Pages
  1--12}.
\newblock Springer~(2011).

\bibitem{aharon2016device}
Nati Aharon, Serge Massar, Stefano Pironio, and Jonathan Silman.
\newblock ``Device-independent bit commitment based on the {CHSH} inequality''.
\newblock
  \href{https://dx.doi.org/https://doi.org/10.1088/1367-2630/18/2/025014}{New
  Journal of Physics {\bf 18}, 025014}~(2016).

\bibitem{aharonov2000quantum4}
Dorit Aharonov, Amnon Ta-Shma, Umesh~V. Vazirani, and Andrew~C. Yao.
\newblock ``Quantum bit escrow''.
\newblock In Proceedings of the Thirty-Second Annual ACM Symposium on Theory of
  Computing.
\newblock \href{https://dx.doi.org/10.1145/335305.335404}{Page 705–714}.
\newblock STOC '00New York, NY, USA~(2000). Association for Computing
  Machinery.

\bibitem{ambainis2004multiparty}
Andris Ambainis, Harry Buhrman, Yevgeniy Dodis, and Hein Rohrig.
\newblock ``Multiparty quantum coin flipping''.
\newblock In Proceedings of the 19th IEEE Annual Conference on Computational
  Complexity.
\newblock
  \href{https://dx.doi.org/https://doi.org/10.1109/CCC.2004.1313848}{Pages
  250--259}.
\newblock ~(2004).

\bibitem{barnum1996noncommuting}
Howard Barnum, Carlton~M Caves, Christopher~A Fuchs, Richard Jozsa, and
  Benjamin Schumacher.
\newblock ``Noncommuting mixed states cannot be broadcast''.
\newblock
  \href{https://dx.doi.org/https://doi.org/10.1103/PhysRevLett.76.2818}{Physical
  Review Letters {\bf 76}, 2818}~(1996).

\bibitem{blum1983coin}
Manuel Blum.
\newblock ``Coin flipping by telephone a protocol for solving impossible
  problems''.
\newblock \href{https://dx.doi.org/https://doi.org/10.1145/1008908.1008911}{ACM
  SIGACT News {\bf 15}, 23--27}~(1983).

\bibitem{chiribella2013short}
Giulio Chiribella, Giacomo~Mauro D{'}Ariano, Paolo Perinotti, Dirk
  Schlingemann, and Reinhard Werner.
\newblock ``A short impossibility proof of quantum bit commitment''.
\newblock
  \href{https://dx.doi.org/https://doi.org/10.1016/j.physleta.2013.02.045}{Physics
  Letters A {\bf 377}, 1076--1087}~(2013).

\bibitem{damgaard2008cryptography}
Ivan~B. Damg\r{a}rd, Serge Fehr, Louis Salvail, and Christian Schaffner.
\newblock ``Cryptography in the bounded-quantum-storage model''.
\newblock \href{https://dx.doi.org/https://doi.org/10.1137/060651343}{SIAM
  Journal on Computing {\bf 37}, 1865--1890}~(2008).

\bibitem{gutoski2007toward}
Gus Gutoski and John Watrous.
\newblock ``Toward a general theory of quantum games''.
\newblock In Proceedings of the Thirty-Ninth Annual ACM Symposium on Theory of
  Computing.
\newblock \href{https://dx.doi.org/10.1145/1250790.1250873}{Page 565–574}.
\newblock STOC '07New York, NY, USA~(2007). Association for Computing
  Machinery.

\bibitem{gutoski2018fidelity}
Gus Gutoski, Ansis Rosmanis, and Jamie Sikora.
\newblock ``Fidelity of quantum strategies with applications to cryptography''.
\newblock
  \href{https://dx.doi.org/https://doi.org/10.22331/q-2018-09-03-89}{Quantum
  {\bf 2}, 89}~(2018).

\bibitem{kilian1988founding}
Joe Kilian.
\newblock ``Founding crytpography on oblivious transfer''.
\newblock In Proceedings of the Twentieth Annual ACM Symposium on Theory of
  Computing.
\newblock \href{https://dx.doi.org/10.1145/62212.62215}{Page 20–31}.
\newblock STOC '88New York, NY, USA~(1988). Association for Computing
  Machinery.

\bibitem{kundu2022device}
Srijita Kundu, Jamie Sikora, and Ernest Y-Z Tan.
\newblock ``A device-independent protocol for {XOR} oblivious transfer''.
\newblock
  \href{https://dx.doi.org/https://doi.org/10.22331/q-2022-05-30-725}{Quantum
  {\bf 6}, 725}~(2022).

\bibitem{lo1997insecurity}
Hoi-Kwong Lo.
\newblock ``Insecurity of quantum secure computations''.
\newblock
  \href{https://dx.doi.org/https://doi.org/10.1103/PhysRevA.56.1154}{Physical
  Review A {\bf 56}, 1154}~(1997).

\bibitem{lo1998quantum}
Hoi-Kwong Lo and Hoi~Fung Chau.
\newblock ``Why quantum bit commitment and ideal quantum coin tossing are
  impossible''.
\newblock
  \href{https://dx.doi.org/https://doi.org/10.1016/S0167-2789(98)00053-0}{Physica
  D: Nonlinear Phenomena {\bf 120}, 177--187}~(1998).

\bibitem{nayak2015quantum}
Ashwin Nayak, Jamie Sikora, and Levent Tunçel.
\newblock ``Quantum and classical coin-flipping protocols based on
  bit-commitment and their point games''~(2015).
\newblock  \href{http://arxiv.org/abs/1504.04217}{arXiv:1504.04217}.

\bibitem{nayak2016search}
Ashwin Nayak, Jamie Sikora, and Levent Tun{\c{c}}el.
\newblock ``A search for quantum coin-flipping protocols using optimization
  techniques''.
\newblock
  \href{https://dx.doi.org/https://doi.org/10.1007/s10107-015-0909-y}{Mathematical
  Programming {\bf 156}, 581--613}~(2016).

\bibitem{rockafellar1970convex}
R~Tyrrell Rockafellar.
\newblock ``Convex {A}nalysis''.
\newblock
  \href{https://dx.doi.org/https://doi.org/10.1515/9781400873173}{Volume~18}.
\newblock Princeton University Press. ~(1970).

\bibitem{schaffner2009robust}
Christian Schaffner, Barbara Terhal, and Stephanie Wehner.
\newblock ``Robust cryptography in the noisy-quantum-storage model''.
\newblock
  \href{https://dx.doi.org/https://doi.org/10.26421/QIC9.11-12-4}{Quantum
  Information and Computation {\bf 9}, 963--996}~(2009).

\bibitem{sikora2014strong}
Jamie Sikora, Andr{\'e} Chailloux, and Iordanis Kerenidis.
\newblock ``Strong connections between quantum encodings, nonlocality, and
  quantum cryptography''.
\newblock
  \href{https://dx.doi.org/https://doi.org/10.1103/PhysRevA.89.022334}{Physical
  Review A {\bf 89}, 022334}~(2014).

\bibitem{sikora2017dierolling}
Jamie Sikora.
\newblock ``Simple, near-optimal quantum protocols for die-rolling''.
\newblock
  \href{https://dx.doi.org/https://doi.org/10.3390/cryptography1020011}{Cryptography
  {\bf 1}, 11}~(2017).

\bibitem{sikora2020impossibility}
Jamie Sikora and John~H Selby.
\newblock ``Impossibility of coin flipping in generalized probabilistic
  theories via discretizations of semi-infinite programs''.
\newblock
  \href{https://dx.doi.org/https://doi.org/10.1103/PhysRevResearch.2.043128}{Physical
  Review Research {\bf 2}, 043128}~(2020).

\bibitem{silman2011fully}
Jonathan Silman, Andr{\'e} Chailloux, Nati Aharon, Iordanis Kerenidis, Stefano
  Pironio, and Serge Massar.
\newblock ``Fully distrustful quantum bit commitment and coin flipping''.
\newblock
  \href{https://dx.doi.org/https://doi.org/10.1103/PhysRevLett.106.220501}{Physical
  Review Letters {\bf 106}, 220501}~(2011).

\bibitem{spekkens2001degrees}
Robert~W Spekkens and Terry Rudolph.
\newblock ``Degrees of concealment and bindingness in quantum bit commitment
  protocols''.
\newblock
  \href{https://dx.doi.org/https://doi.org/10.1103/PhysRevA.65.012310}{Physical
  Review A {\bf 65}, 012310}~(2001).

\bibitem{wehner2008cryptography}
Stephanie Wehner, Christian Schaffner, and Barbara~M Terhal.
\newblock ``Cryptography from noisy storage''.
\newblock
  \href{https://dx.doi.org/https://doi.org/10.1103/PhysRevLett.100.220502}{Physical
  Review Letters {\bf 100}, 220502}~(2008).

\bibitem{yao1986generate}
Andrew Chi-Chih Yao.
\newblock ``How to generate and exchange secrets''.
\newblock In 27th Annual Symposium on Foundations of Computer Science (sfcs
  1986).
\newblock \href{https://dx.doi.org/https://doi.org/10.1109/SFCS.1986.25}{Pages
  162--167}.
\newblock ~(1986).

\end{thebibliography}


\appendix 


\section{An example where the stochastic switch does not help}\label{appendix:xot:dieroll}

We now illustrate an example where the stochastic switch by Bob does not provide any improvement in the overall security of a cheating Alice. The protocol we detail switches between XOR oblivious transfer and die rolling with three possible outcomes.

\subsection{XOR oblivious transfer}\label{subsection:obliviousTransfer:XOR}
XOR oblivious transfer (XOT) is a cryptographic task between Alice and Bob who wish to achieve the following. Alice outputs a trit $y \in \{0,1,2\}$ while Bob outputs two bits $(x_0,x_1)$ each by a uniformly random sampling, and they both communicate in a way that Bob learns $x_y$ by the end of the protocol (here $x_2$ is defined as $x_0 \oplus x_1$). 

\begin{itemize}
    \item \emph{Completeness:} An XOT protocol is said to be \emph{complete} if Alice learns the selected bit ($x_y$) unambiguously.
    \item \emph{Cheating Alice:} Dishonest Alice could deviate from an XOT protocol in order to learn both bits $(x_0,x_1)$ with the corresponding cheating probability given by
    \begin{equation*}
        P_A^{XOT} = \Pr[\text{Alice correctly guesses $(x_0,x_1)$}].
    \end{equation*}
    \item \emph{Cheating Bob:} Dishonest Bob on the other hand would attempt to learn Alice's choice trit $y$ with the corresponding cheating probability given by 
    \begin{equation*}
        P_B^{XOT} = \Pr[\text{Bob correctly guesses $y$}].
    \end{equation*}
\end{itemize}

A simple extension of the $1$-out-of-$2$ OT Protocol (\cref{subsection:obliviousTransfer:1outOf2}) for the XOT task is provided in \cite{kundu2022device} and is reproduced shortly.

The following three two-qutrit states will be used throughout this section. 

Let $\ket{\phi_y}$ be the following two-qutrit state in $\A \otimes \B$:
\begin{equation}\label{states:obliviousTransfer:XOT}
    \ket{\phi_y} 
    = 
    \begin{cases} 
    \quad \frac{1}{\sqrt{2}}(\ket{00}_{\A\B} + \ket{22}_{\A\B}) & \text{ if } y = 0, \\ 
    \quad \frac{1}{\sqrt{2}}(\ket{11}_{\A\B} + \ket{22}_{\A\B}) & \text{ if } y = 1, \\ 
    \quad \frac{1}{\sqrt{2}}(\ket{00}_{\A\B} + \ket{11}_{\A\B}) & \text{ if } y = 2. 
    \end{cases} 
    \end{equation}   
\begin{protocol}{Quantum XOR oblivious transfer \cite{kundu2022device}.}\label{protocol:obliviousTransfer:XOR}
    \begin{itemize}
        \underline{Stage-I}
        \item \textbf{Alice chooses a trit $y \in \{0,1,2\}$ uniformly at random and creates the two-qutrit state $\ket{\phi_y}$ as defined in \cref{states:obliviousTransfer:XOT}, above.
        \item Alice sends the qutrit $\mathcal{B}$ to Bob.}\par
        \underline{Stage-II}
        \item Bob selects $(x_0, x_1) \in \{0,1\}^2$ uniformly at random and applies the unitary 
        \begin{equation*}
            U_{x_0x_1} = 
            \begin{bmatrix}
                (-1)^{x_0} & 0 & 0 \\
                0 & (-1)^{x_1} & 0 \\
                0 & 0 & 1
            \end{bmatrix}
        \end{equation*}
        to the received qutrit. Afterwards, Bob returns the qutrit to Alice.
        \item Alice determines the value of $x_y$ using the two-outcome measurement:
        \begin{equation*}
            \{\Pi_{0} \coloneqq \kb{\phi_y}_{\A \otimes \B}, \Pi_{1} \coloneqq \mathbbm{1}_{\A \otimes \B} - \kb{\phi_y}_{\A \otimes \B} \}.
        \end{equation*}
    \end{itemize}
\end{protocol}

Note that \cref{protocol:obliviousTransfer:XOR} above has a significant overlap with \cref{protocol:obliviousTransfer:1outOf2} for OT as Alice wants to learn $(x_0,x_1)$ in both protocols. The only difference is the state created by Alice at the beginning of the two protocols. As the sequence and actions of Bob on the qutrit $\B$ remains same in both the protocols, we may infer that the optimal cheating strategy for dishonest Alice is the same in both Protocols \ref{protocol:obliviousTransfer:XOR} and \ref{protocol:obliviousTransfer:1outOf2} implying $P_A^{XOT} = 3/4$ achieved by Alice by sending the qutrit $\B$ at the beginning in the state
$\begin{bmatrix}
        1/3 & 0 & 0 \\
        0 & 1/3 & 0 \\
        0 & 0 & 1/3
\end{bmatrix}$.

\subsection{Die rolling}\label{subsection:dieRolling}
Die rolling (DR) is the cryptographic task between two parties (Alice and Bob) where they communicate to agree upon a common integer outcome from the set $\{1,2,\ldots D\}$. The security notions of a die rolling protocol are defined below.
\begin{itemize}
    \item \emph{Completeness:} If both Alice and Bob are honest, then neither party aborts and they obtain the same outcome uniformly at random.
    \item \emph{Cheating Alice:} Dishonest Alice could deviate from the protocol to force an outcome $d$, the extent of which is given by
    \begin{equation*}
        P_{A,d}^{DR} = \max \Pr[\text{Alice successfully forces outcome $d$}]
    \end{equation*}
    where the maximum is taken over all strategies of Alice. Note that there could be different cheating probabilities for each $d \in \{1,2,\ldots D\}$.
    \item \emph{Cheating Bob:} A dishonest Bob could attempt to force an outcome $d \in \{1,2,\ldots D\}$ the extent of which is given by
    \begin{equation*}
        P_{B,d}^{DR} = \max \Pr[\text{Bob successfully forces outcome $d$}]
    \end{equation*}
    where the maximum is taken over all strategies of Bob. As earlier, $P_{B,d}^{DR}$ could be different for each $d \in \{1,2,\ldots D\}$. 
\end{itemize}
For the task of die rolling with three possible outcomes, we consider the following protocol.

\begin{protocol}{Quantum die rolling with three outcomes ($D = 3$)~\cite{sikora2017dierolling}.}\label{protocol:dieRoll:integerCommitment}
    \begin{itemize}
        \underline{Stage-I}
        \item \textbf{Alice chooses a trit $y \in \{0,1,2\}$ uniformly at random and creates the two-qutrit state $\ket{\phi_y}$ as defined in \cref{states:obliviousTransfer:XOT}.
        \item Alice sends the qutrit $\mathcal{B}$ to Bob.}\par
        \underline{Stage-II}
        \item Bob selects a uniformly random $z \in \{0,1,2\}$ and sends it to Alice.
        \item Alice reveals $y$ to Bob and sends him the qutrit $\A$.
        \item Bob measures the combined state $\A \otimes \B$ to accept or reject with the POVM:
        \begin{equation*}
            \{\Pi_{accept} \coloneqq \kb{\phi_y}_{\A \otimes \B}, \Pi_{reject} \coloneqq \mathbbm{1}_{\A \otimes \B} - \Pi_{accept}\}.
        \end{equation*}
        If Bob accepts, Alice and Bob output $d \coloneqq (y+z\mod{3})+1$ as the outcome of the protocol.
    \end{itemize}
\end{protocol}

For \cref{protocol:dieRoll:integerCommitment}, the uniformly random selection of $y$ and $z$ indicates that $P_{A,0}^{DR} = P_{A,1}^{DR} = P_{A,2}^{DR} \coloneqq P_{A}^{DR}$ and $P_{B,0}^{DR} = P_{B,1}^{DR} = P_{B,2}^{DR} \coloneqq P_{B}^{DR}$. The SDP formulation for cheating Alice in \cref{protocol:dieRoll:integerCommitment} extends trivially from its SDP formulation in \cref{protocol:bitCommitment}. Solving it numerically, we find that Alice can cheat with an optimal probability of $P_A^{DR} = 2/3$ achieved when Alice sends the qutrit $\B$ at the beginning in the state 
$\begin{bmatrix}
    1/3 & 0 & 0 \\
    0 & 1/3 & 0 \\
    0 & 0 & 1/3
\end{bmatrix}$.

The Stage-I communication between Alice and Bob is common to both the Protocols \ref{protocol:obliviousTransfer:XOR} and \ref{protocol:dieRoll:integerCommitment} and hence we are able to develop a switch protocol between XOT and DR as detailed below.
\begin{protocol}{Quantum XOT-DR stochastic switch.}\label{protocol:switch:obliviousTransfer:XOR:dieRoll:integerCommitment}
    \begin{itemize}
        \underline{Stage-I}
        \item \textbf{Alice chooses a trit $y \in \{0,1,2\}$ uniformly at random and creates the two-qutrit state $\ket{\phi_y}$ as defined in \cref{states:obliviousTransfer:XOT}.
        \item Alice sends the qutrit $\mathcal{B}$ to Bob.}\par
        \underline{Stage-II}
        \item Bob chooses $c \in \{0,1\}$ uniformly at random and sends it to Alice.
        \item If $c = 0$, Alice and Bob perform XOR oblivious transfer as per \cref{protocol:obliviousTransfer:XOR}, otherwise, they perform the three-outcome die rolling \cref{protocol:dieRoll:integerCommitment}.  
    \end{itemize}
\end{protocol}

\textbf{Cheating Alice.} Dishonest Alice in \cref{protocol:switch:obliviousTransfer:XOR:dieRoll:integerCommitment} would like to send a first message such that her average success of correctly guessing $(x_0,x_1)$ and forcing an outcome is maximized. It is interesting to note that the optimal message from Alice can be the same in both \cref{protocol:obliviousTransfer:XOR} and \cref{protocol:dieRoll:integerCommitment} implying that the optimal cheating probability of Alice in \cref{protocol:switch:obliviousTransfer:XOR:dieRoll:integerCommitment} is simply the average of the optimal cheating probabilities $P_A^{XOT}$ and $P_A^{DR}$. The identical nature of the optimal first message in the individual protocols indicates that the stochastic switch by Bob to introduce uncertainty on Alice's side is not improving the security against cheating Alice in the switch version. We note again that if optimal sets of first messages of two or more protocols have a non-empty intersection, then it is not possible to take advantage of the switching paradigm as illustrated in \cref{fig:optimalSets}.


\section{An example where Bob can take advantage of stochastic switch}\label{appendix:bitCommitment:EPR}

Next we illustrate an example where the stochastic switch by Bob breaks the overall security of the protocol. We describe a switch protocol for strong coin flipping based on bit commitment (\cref{subsection:bitCommitment}) and weak coin flipping (\cref{subsection:coinFlip:EPR}).  

Strong coin flipping is the cryptographic task between Alice and Bob where they communicate to agree on a common binary outcome. The security notions of strong coin flipping (SCF) are stronger (hence the name) than weak coin flipping (WCF) because in SCF, Alice and Bob each may try to bias the outcome towards either $0$ or $1$. The SCF task can also be viewed as a die rolling task with only two possible outcomes (See \cref{subsection:dieRolling}).

Consider a two-outcome variant of the die rolling protocol discussed previously in \cref{protocol:dieRoll:integerCommitment} and outlined below.
\begin{protocol}{Quantum strong coin flipping \cite{kerenidis2004weak}.}\label{protocol:SCF:bitCommitment}
    \begin{itemize}
        \underline{Stage-I}
        \item \textbf{Alice chooses a bit $y \in \{0,1\}$ uniformly at random and creates the two-qutrit state}
        \begin{equation*}
            \ket{\phi_y} = \frac{1}{\sqrt{2}}\ket{yy} + \frac{1}{\sqrt{2}}\ket{22} \in \A \otimes \B.
        \end{equation*}
        \item \textbf{Alice sends the qutrit $\mathcal{B}$ to Bob.}\par
        \underline{Stage-II}
        \item Bob selects a uniformly random $z \in \{0,1\}$ and sends it to Alice.
        \item Alice reveals $y$ to Bob and sends him the qutrit $\A$.
        \item Bob measures the combined state $\A \otimes \B$ to accept or reject with the POVM:
        \begin{equation*}
            \{\Pi_{accept} \coloneqq \kb{\phi_y}, \Pi_{reject} \coloneqq \mathbbm{1}_{\A \otimes \B} - \Pi_{accept}\}.
        \end{equation*}
        If Bob accepts, Alice and Bob output $y \oplus z$ as the outcome of the protocol.
    \end{itemize}
\end{protocol}

Note that the WCF protocol given by \cref{protocol:coinFlip:EPR} is a valid SCF protocol as well since the final outcome is a uniformly random bit when both Alice and Bob are honest. A new SCF protocol based on stochastic switch between \cref{protocol:SCF:bitCommitment} and \cref{protocol:coinFlip:EPR} is given next.
\newpage
\begin{protocol}{Quantum SCF based on DR-EPR stochastic switch.}\label{protocol:switch:SCF:bitCommitment:coinFlip:EPR}
    \begin{itemize}
        \underline{Stage-I}
        \item \textbf{Alice chooses a bit $y \in \{0,1\}$ uniformly at random and creates the two-qutrit state 
        \begin{equation*}
            \ket{\phi_y} = \frac{1}{\sqrt{2}}\ket{yy} + \frac{1}{\sqrt{2}}\ket{22} \in \A \otimes \B.
        \end{equation*}
        \item Alice sends the qutrit $\mathcal{B}$ to Bob.}\par
        \underline{Stage-II}
        \item Bob chooses $c \in \{0,1\}$ uniformly at random and sends it to Alice.
        \item If $c = 0$, Alice and Bob perform SCF as per \cref{protocol:SCF:bitCommitment}, otherwise, they perform SCF as per \cref{protocol:coinFlip:EPR}. 
    \end{itemize}
\end{protocol}

From the analysis of the previous switch protocols, one might presume that the security of \cref{protocol:switch:SCF:bitCommitment:coinFlip:EPR} is at least as good as its constituent protocols. However, it is broken when Bob is dishonest. To see this consider the following strategy by Bob attempting to force the outcome $1$ where he simply measures the received qutrit (at the end of Stage-I) in the computational basis. If the outcome is either $0$ or $1$, he already knows Alice's bit $y$ with certainty and selects $c = 0$ to perform SCF based on bit commitment (\cref{protocol:SCF:bitCommitment}). This observed value of $y$ allows him to successfully force outcome $0$ or $1$ with probability $1$. Alternatively, if the measurement outcome is $2$, he selects $c = 1$ to perform \cref{protocol:coinFlip:EPR}. Note that the state in $\A$ has collapsed to $\ket{2}$. Thus, Bob can use this to force outcome $1$ again with probability $1$. Therefore, Bob can measure and, conditioned on the outcome, cheat in one of the protocols with certainty.


\end{document}